\documentclass[12pt,14paper]{article}
\usepackage[english]{babel}
\usepackage[latin1]{inputenc}
\usepackage[normalem]{ulem}
\usepackage[intlimits]{amsmath}
\usepackage{amssymb}
\usepackage{graphics}
\usepackage{geometry}
\usepackage{color}
\usepackage{verbatim} 
\usepackage{axodraw}
\usepackage{multirow} 
\usepackage{enumitem}

\usepackage{array}
\usepackage{float}
\usepackage{lscape,graphicx} 
\usepackage{amssymb}
\usepackage{advdate}

\newcommand{\GeV}{\,\mbox{GeV}}

\newcommand{\bea}{\begin{equation}\begin{array}{c}}
\newcommand{\eea}{\end{array}\end{equation}}
\newcommand{\ea}{\end{array}}

\newcommand{\beq}{\begin{equation}}
\newcommand{\eeq}{\end{equation}}
\newcommand{\bad}{\begin{array}{ccc}}

\newcommand{\ba}{\begin{array}{c}}

\begin{document}

\hfill{{\small Ref. FLAVOUR(267104)-ERC-57}}

\hfill{{\small Ref. TUM-HEP 911/13}}

\begin{center}
\mathversion{bold}
{\LARGE\bf{  Dark matter production  from Goldstone boson interactions
and implications for direct searches and  dark radiation}}
\mathversion{normal}

\vspace{1cm}
Camilo Garcia-Cely, Alejandro Ibarra and Emiliano Molinaro\\
\vspace{0.5cm}
{\em Physik-Department T30d, Technische Universit\"at M\"unchen,\\ James-Franck-Stra{\ss}e, 85748 Garching, Germany.\\}
\end{center}

\begin{abstract}
The stability of the dark matter particle could be attributed to the remnant $Z_2$ symmetry that arises from the spontaneous breaking of a global $U(1)$ symmetry.
This plausible scenario contains a Goldstone boson which, as recently shown by Weinberg, is a strong candidate for dark radiation. We show in this paper that this Goldstone boson, together with the $CP$-even scalar associated to the spontaneous breaking of the global $U(1)$ symmetry, plays a central role in the dark matter production. Besides, the mixing of the $CP$-even scalar with the Standard Model Higgs boson leads to novel Higgs decay channels and to interactions with nucleons, thus opening the possibility of probing this scenario at the LHC and in direct dark matter search experiments. We carefully analyze the latter possibility and we show that there are good prospects to observe a signal at the future experiments LUX and XENON1T provided the dark matter particle was produced thermally and has a mass larger than $\sim 25$ GeV.
\end{abstract}

\section{Introduction}

Numerous observations support the hypothesis that the 85\% of the matter content of the Universe is in the form of a new particle, the dark matter particle (for reviews see \cite{Bertone:2004pz,Bergstrom:2000pn}). One of the most striking features of this new particle is its long lifetime, longer than the age of the Universe and possibly much longer, as indicated by the non-observation of its decay products in cosmic ray experiments~\cite{Ibarra:2013cra}. 

The longevity of the dark matter particle is very likely due to the existence of a preserved, or very mildly broken, symmetry in the Lagrangian (see \cite{Hambye:2010zb} for a review of possible explanations to the dark matter stability). The simplest symmetry that ensures the absolute stability of the dark matter particle is a discrete $Z_2$ symmetry, under which all the Standard Model particles are even while the dark matter particle (and possibly other particles in the dark sector) are odd. The discrete symmetry in the Lagrangian could be imposed {\it ab initio} or could, perhaps more plausibly, arise as a remnant of the breaking of a global continuous symmetry. Indeed, if a global $U(1)$ symmetry is spontaneously broken by a scalar field with charge 2 under that symmetry, a discrete $Z_2$ symmetry automatically arises in the Lagrangian. Moreover, all the fields with even (odd) charge under the global group will acquire, after the spontaneous symmetry breaking, an even (odd) discrete charge under the $Z_2$ transformation~\cite{Krauss:1988zc}. Therefore, the lightest particle with odd charge is absolutely stable and a potential candidate for dark matter.

The spontaneous breaking of a global continuous  symmetry, as is well known, gives rise to massless Goldstone bosons in the spectrum. While the presence of Goldstone bosons is usually an unwanted feature in model building, it was recently argued by Weinberg that the Goldstone boson that arises in this framework could contribute to the dark radiation. More specifically, if the Goldstone bosons are in thermal equilibrium with the Standard Model particles until the era of muon annihilation, their contribution to the effective number of neutrino species $N_{\rm eff}$ would be 0.39~\cite{Weinberg:2013kea}, in remarkable agreement with the central value obtained in \cite{Planck2013} from combining Planck data, WMAP9 polarization data and ground-based observations of high-$\ell$, which imply $N_{\rm eff}=3.36^{+0.68}_{-0.64}$ at 95\% C.L.

We will argue in this paper that the Goldstone bosons associated to the stability of the dark matter particle are not only a potential candidate for dark radiation, but also play a crucial role in the dark matter production\footnote{Similar ideas have been discussed in the context of Majoron models \cite{Lindner:2011it, JosseMichaux:2011ba}, where the global $U(1)$ symmetry corresponds to the $B-L$ charge.}.~In fact, in most of the parameter space, the most important processes for dark matter production are the (co)annihilations into the scalar particles in the dark sector responsible for the spontaneous breaking of the global $U(1)$ symmetry, namely a $CP$-even scalar and a Goldstone boson. Moreover, the $CP$-even scalar in general mixes with the Standard Model Higgs boson, possibly producing deviations from the Standard Model prediction in the invisible decay width of the Higgs \cite{Weinberg:2013kea}, which in this framework could also decay into particles of the dark sector. The latter could give concrete signatures which can be searched for at the LHC \cite{Cheung:2013oya}. 
Interestingly, the mixing of the $CP$-even scalar of the dark sector with the Standard Model Higgs boson leads to the interaction of the dark matter particle with nucleons, thus opening the possibility of detecting signatures of this model also in direct dark matter search experiments. We will see that this is indeed the case and that  XENON100 \cite{Aprile:2012nq} - and recently the LUX experiment \cite{Akerib:2013tjd} - already probes parts of the relevant parameter space of the model, in particular regions where Goldstone bosons sizably contribute to the number of neutrino species before recombination. 
See \cite{Anchordoqui:2013pta} for related analyses.

The paper is organized as follows. In section \ref{sec:model} we describe the scalar and dark matter sectors of the model as well as the existing limits on the model parameters from invisible Higgs decays. Then, in section \ref{sec:relic_abundance} we calculate the relic abundance of dark matter particles and in section \ref{sec:direct_detection} the limits on the model parameters from the  LUX experiment. In section \ref{sec:dark_radiation} we revisit the conditions under which the Goldstone bosons decouple from the thermal bath just before the era of muon annihilation, in order to contribute to the effective number of degrees of freedom as $\Delta N_{\rm eff}\simeq 0.39$. In section \ref{sec:exclusionplots} we show that the explanation of the dark radiation of the Universe in terms of Goldstone bosons can be probed in direct dark matter search experiments. We show that the XENON100 and the LUX experiments already rule out parts of the parameter space of the model, and we argue that there are good prospects to observe a signal at  LUX (final phase) or XENON1T if the dark matter particles were thermally produced. Lastly, in section \ref{sec:conclusions} we present our conclusions.

\section{Description of the Model}
\label{sec:model}

We consider the model proposed in \cite{Weinberg:2013kea}, where the Standard Model (SM) is extended by one complex scalar field $\phi$  and one Dirac fermion field $\psi$. The new fields are singlets under the SM gauge group and are charged under a global $U(1)_\text{DM}$ symmetry, namely: $U(1)_\text{DM}(\psi)=1$ and $U(1)_\text{DM}(\phi)=2$. 
Besides, all the SM fields transform trivially under the global symmetry. As a consequence, the dark sector interacts with the SM fermions only through the Higgs portal. The corresponding interaction Lagrangian reads:
\begin{eqnarray}
	\mathcal{L} & =& \left(D_{\mu}H\right)^{\dagger}\left( D^{\mu}H\right)\,+ \,\mu_{H}^{2}\,H^{\dagger}\,H\,-\,\lambda_{H}\,\left(H^{\dagger}\,H\right)^{2}\nonumber\\
	&&\,+\,\partial_{\mu}\phi^{*} \partial^{\mu}\phi\,+\,	\mu_{\phi}^{2}\,\phi^{*}\,\phi\,-\,\lambda_{\phi}\,\left(\phi^{*}\,\phi\right)^{2}\,-\,
	\kappa\,\left(H^{\dagger}\,H\right)\,\left(\phi^{*}\,\phi\right)+\mathcal{L_{\text{DM}}}\;,
\label{lint}
\end{eqnarray}
where $D_{\mu}$ is the covariant derivative, $H$ is the SM Higgs doublet and
\begin{equation}
{\cal L_\text{DM}} = i \overline{\psi}\gamma^\mu \partial_\mu \psi -M\overline{\psi}\psi - \left(\dfrac{f}{\sqrt{2}}\phi \overline{\psi} \psi^c + h.c.\right)\,,
\label{DiracL}
\end{equation}
with $\psi^{c}\equiv C \overline{\psi}^{T}$. In the following we discuss separately the scalar sector and the dark matter sector.
\subsection{The Scalar Sector}
Both the scalar field $\phi$ and the neutral component of the Higgs doublet acquire non-zero vacuum expectation values (vev), which spontaneously break the symmetry group $SU(3)_{\rm c}\times SU(2)_{\rm W}\times U(1)_{\rm Y}\times [U(1)_\text{DM}] \to U(1)_{\rm em} \times Z_{2} $. In order to analyze the physical mass spectrum of the theory, we conveniently parametrize the scalar fields in Eq.~(\ref{lint}) as:~\footnote{In contrast to \cite{Weinberg:2013kea}, with this parametrization only renormalizable terms in the Lagrangian are necessary to analyze, at lowest order, the phenomenology of the dark sector.}

\begin{equation}
H  =\begin{pmatrix} G^+ \\ \frac{v_H+\tilde{h}+i G^0 }{\sqrt{2}} \end{pmatrix} \;, \hspace{40pt} \phi  = \frac{v_\phi+\tilde{\rho}+i\eta}{\sqrt{2}} \;,
\label{fieldHphi}
\end{equation}
where the SM Higgs vev is $v_{H}\simeq 246$ GeV. Then, from the minimization of the scalar potential in Eq.~(\ref{lint}) we get the following tree-level relations between the parameters of the Lagrangian and the vacuum expectation values:
\begin{eqnarray}
	\mu_{H}^{2} = \frac{1}{2}\left(2\,v_{H}^{2}\,\lambda_{H}\,+\,v_{\phi}^{2}\,\kappa  \right)\,,\hspace{40pt}
	\mu_{\phi}^{2} = \frac{1}{2}\left(2\,v_{\phi}^{2}\,\lambda_{\phi}\,+\,v_{H}^{2}\,\kappa  \right) \,.
\label{eq:mass_parameters}
\end{eqnarray}
The neutral $CP$-odd component of the Higgs doublet, $G^0$, provides the longitudinal polarization of the $Z$ boson through the Brout-Englert-Higgs mechanism. On the other hand, the pseudo-scalar field $\eta$ corresponds to the Goldstone boson that arises from the spontaneous breaking of the global $U(1)_\text{DM}$ symmetry. Therefore, the physical mass spectrum  consists of two $CP$-even massive real scalars, denoted by $h$ and $\rho$, which are linear combinations of the interaction fields $\tilde{h}$ and $\tilde{\rho}$ in Eq.~(\ref{fieldHphi}), and a $CP$-odd massless scalar $\eta$.
The mass matrix of the $CP$-even scalars in the basis of interaction fields $(\tilde{h},\,\tilde{\rho})$ reads
\begin{eqnarray}
	\mathcal{M}_{S}\;=\;
\left(
\begin{array}{cc}
2\,\lambda_{H} \,v_{H}^{2}& \kappa\,v_{H}\,v_{\phi}  \\
\kappa\,v_{H}\,v_{\phi} & 2\,\lambda_{\phi}\,v_{\phi}^{2}
\end{array}
\right)\,,
\end{eqnarray}
where we have used Eq.~(\ref{eq:mass_parameters}). The mass eigenstates $h$ and $\rho$ are thus obtained by the basis transformation:
\bea
\left(\ba \tilde{h}\\ \tilde{\rho} \ea\right)=R_{S}\,\left(\ba h\\\rho\ea\right)\,,
\eea
with 
\bea
	R_{S} \;\equiv\;\left( \begin{array}{cc}
\cos\theta& \sin\theta \\
-\sin\theta& \cos\theta
\end{array}
\right)\quad\text{and}\quad \tan2\theta\;=\; \frac{\kappa \,v_{H}\,v_{\phi}}{\lambda_{\phi}\,v_{\phi}^{2}\,-\,\lambda_{H}\,v_{H}^{2}}\,.\label{kappa}
\eea
The masses of the physical states are:
\begin{eqnarray}
 m_{h}^{2} & = & 
 2\,\lambda_{H}\,v_{H}^{2}\,\cos^{2}\theta\,+\,2\,\lambda_{\phi}\,v_{\phi}^{2}\,\sin^{2}\theta \label{mass1}
 \,-\,\kappa\,v_{H}\,v_{\phi}\,\sin2\theta\,,\\
 m_{\rho}^{2} & = & 
 2\,\lambda_{H}\,v_{H}^{2}\,\sin^{2}\theta\,+\,2\,\lambda_{\phi}\,v_{\phi}^{2}\,\cos^{2}\theta \label{mass2}
 \,+\,\kappa\,v_{H}\,v_{\phi}\,\sin2\theta\,.
\end{eqnarray}
In the following, we assume that the $CP$-even state $h$ corresponds to the Standard Model Higgs boson with $m_{h}=125$ GeV. Notice that, after the electroweak symmetry breaking, the scalar sector can be described in terms of three independent unknown parameters, $m_{\rho}$, $v_{\phi}$ and $\theta$, as well as the mass and the vev of the Higgs. With this choice, the quartic couplings are unambiguously given by 
\begin{eqnarray}\label{kappaeq}
\lambda_H  = \frac{m_h^2 \cos^2\theta + m_\rho^2 \sin^2\theta}{2 v_H^2},\hspace{30pt}
\lambda_\phi  = \frac{m_h^2 \sin^2\theta + m_\rho^2 \cos^2\theta}{2 v_\phi^2},&&\\
\kappa =\frac{(m_{\rho}^{2}-m_{h}^{2})\,\sin2\theta}{2\,v_{H}\,v_{\phi}}\,.\hspace{100pt} &&\nonumber
\end{eqnarray}
The stability of the scalar potential implies the condition $4\, \lambda_{H}\,\lambda_{\phi}\,-\,\kappa^{2}>0$, which is automatically satisfied by the previous equations as long as $m_\rho^2>0$ and $m_h^2>0$.

\subsection{The Dark Matter Sector}

The coupling constant $f$ of the interaction between the Dirac field $\psi$ and the complex scalar $\phi$ in Eq.~(\ref{DiracL}) is in general complex. However, this phase can be absorbed by a redefinition of the scalar field $\phi$. As a result, the Lagrangian Eq.~(\ref{DiracL}) conserves $CP$ and both $P$ and $C$ separately. Besides, the Dirac field $\psi$ is no longer a mass eigenstate after $H$ and $\phi$ acquire non-zero vacuum expectation values. Indeed it splits into two new mass-eigenstates, which correspond to the Majorana fermions:
\begin{equation}
\psi_+ = \frac{\psi+\psi^c}{\sqrt{2}}, \hspace{40pt} \psi_- = \frac{\psi-\psi^c}{\sqrt{2}i}\,,
\end{equation}
which are  $CP$-even and $CP$-odd respectively. In terms of them, the Lagrangian can be cast as 
\begin{eqnarray}
{\cal L} &=& \frac{1}{2} \left(i \overline{\psi_+}\gamma^\mu \partial_\mu \psi_+ + i\overline{\psi_-}\gamma^\mu \partial_\mu \psi_- -M_+ \overline{\psi_+} \psi_+  -  M_-\overline{\psi_-} \psi_- \right) \nonumber \\
&&-\, \dfrac{f}{2} \left((-\sin\theta\, h + \cos\theta\, \rho)(\overline{\psi_+} \psi_+ - \overline{\psi_-} \psi_-) +\eta\,(\overline{\psi_+} \psi_- + \overline{\psi_-} \psi_+)  \right)\,,
\label{DiracLext}
\end{eqnarray}
with $M_\pm =|M\pm f v_\phi|$. Notice that this Lagrangian is invariant under the $Z_{2}$ transformation $\psi_{\pm}\to - \psi_{\pm}$, which is a remnant of the spontaneously broken $U(1)_{\text{DM}}$ symmetry. As a result, the lightest Majorana fermion is stable and, consequently, a dark matter candidate. If the coupling constant $f$ is positive, the lightest Majorana fermion is $\psi_-$. Without loss of generality we will assume that this is the case. Notice that the dark sector contains five unknown parameters, for example, $M_{-}$, $m_{\rho}$, $\lambda_{\phi}$, $\theta$ and $f$. Nevertheless, in some instances we will find convenient to express observables in terms of the following dimensionless quantities:
\begin{equation}
r = \frac{m_\rho}{M_-}, \hspace{40pt} z = \frac{M_+}{M_-}\,.
\end{equation}

In this scenario, not only the dark matter particle survives until today, but also the Goldstone boson. \footnote{If the global symmetry is explicitly broken, the Goldstone boson could have a small mass and decay into two photons. We will, however, not consider this possibility in this paper.} In fact, all other particles of the dark sector are unstable. On the one hand, $\psi_{+}$ decays into a dark matter particle and a Goldstone boson with a decay rate:
\begin{equation}
\Gamma(\psi_+ \to \psi_-\eta)=\frac{f^2 (M_{+}^{2}-M_{-}^{2})(M_{+}+M_{-})^{2}}{16 \pi \,M_{+}^{3}} \,.
\end{equation}
On the other hand, the scalar $\rho$ decays into $\psi^{\pm}$ pairs, two Goldstone bosons or SM particles. Decays into SM particles are negligible since the corresponding decay rate is proportional to $\sin^{2}\theta$, which, as we will see in the next subsection, should be very small. Therefore the relevant decay widths read:
\begin{eqnarray}
	&&\Gamma(\rho\to \eta\,\eta)  \;= \; \frac{m_{\rho}^{3}\,\cos^{2}\theta}{32\,\pi\,v_{\phi}^{2}},\label{r0invdec}\\
	&&\Gamma\left(\rho\to \psi_\pm\psi_\pm\right) \; = \;\frac{f^{2}\,\cos^{2}\theta}{16\,\pi\,m_{\rho}^{2}}\, \left(m_{\rho}^{2}-4\,M_{\pm}^2\right)^{3/2}\,.\label{rtoDMDM}
\end{eqnarray}

\subsection{Constraints from Invisible Higgs Decays}

The enlarged scalar and fermion sectors affect the SM Higgs decay channels. The new decay modes and the corresponding decay rates are:
\begin{eqnarray}
	&&\Gamma(h\to \eta\,\eta)  \;= \; \frac{m_{h}^{3}\,\sin^{2}\theta}{32\,\pi\,v_{\phi}^{2}},\label{H0invdec}\\
	&&\Gamma(h\to \rho \,\rho) \; = \;\frac{\left(m_{h}^{2}+2\,m_{\rho}^{2}\right)^{2}}{128\,\pi\,m_{h}^{2}\,v_{H}^{2}\,v_{\phi}^{2}}\, \sqrt{m_{h}^{2}-4m_{\rho}^{2}}\,\left(v_{H}\cos\theta-v_{\phi}\sin\theta\right)^{2}\sin^{2}2\theta,\label{htorhorho}\\
	&&\Gamma\left(h\to \psi_\pm\psi_\pm\right) \; = \;\frac{f^{2}\,\sin^{2}\theta}{16\,\pi\,m_{h}^{2}}\, \left(m_{h}^{2}-4\,M_{\pm}^2\right)^{3/2}\,.\label{htoDMDM}
\end{eqnarray}
It is possible to constrain the value of the scalar mixing angle $\theta$ from the experimental upper bound on the Higgs boson invisible decay width. Indeed, neglecting for simplicity the $h$ decays into a pair of $\rho$ or $\psi_{\pm}$, the total decay width of $h$ takes the form:
\begin{equation}
	\Gamma_{h}^{\rm tot} \;=\; \cos^{2}\theta\,\Gamma_{\rm Higgs}^{\rm SM}\,+\,\Gamma\left(h\to\eta\,\eta\right)\,,\label{Gtot}
\end{equation}
where $\Gamma_{\rm Higgs}^{\rm SM}\simeq 4$ MeV is the total decay width of the Higgs boson within the Standard Model for a Higgs boson mass of $125\GeV$. Therefore, from Eq.~(\ref{Gtot}) it follows that
\begin{equation}
	\Gamma\left(h\to\eta\,\eta\right)\;<\; \frac{B_{\rm inv}\,\cos^{2}\theta}{1-B_{\rm inv}}\,\Gamma_{\rm Higgs}^{\rm SM}\,,
\end{equation}
where $B_{\rm inv}\simeq 20\%$ (see, $e.g.$, \cite{Belanger:2013kya,Giardino:2013bma})  is the conservative experimental upper limit on the invisible branching ratio of the Higgs boson. Thus, from the expression of the $h$ decay rate into two Goldstone bosons, Eq. (\ref{H0invdec}), the following upper limit on $\tan\theta$ can be derived~\cite{Weinberg:2013kea}:
\begin{equation}
	\left|\tan\theta\right| <  \sqrt{\frac{32\,\pi\,v_{\phi}^{2}\,\Gamma_{\rm Higgs}^{\rm SM}\,B_{\rm inv}}{m_{h}^{3}\,(1-B_{\rm inv})}}
	\;\lesssim\;2.2\times 10^{-3}\,\left(\frac{v_{\phi}}{10\,{\rm GeV}}\right)\approx1.6\times 10^{-5}\lambda_{\phi}^{-1/2}\left(\frac{m_{\rho}}{0.1\,{\rm GeV}}\right)\,,
\label{thetabound}
\end{equation}
where in the last expression it was replaced $v_\phi\approx m_{\rho}/\sqrt{2\,\lambda_{\phi}}$ at leading order in $\theta$.

Including the other two decay processes, Eqs.~(\ref{htorhorho}) and (\ref{htoDMDM}), when kinematically allowed, would reduce the upper bound derived in Eq.~(\ref{thetabound}) by up to 5\%. As we will see in Section 4, stronger limits on the scalar mixing angle $\theta$ can be derived from dark matter direct detection experiments.

\section{Dark Matter Relic Abundance}
\label{sec:relic_abundance}

\begin{figure}[t!]
\begin{center}
\includegraphics[width=0.7\textwidth]{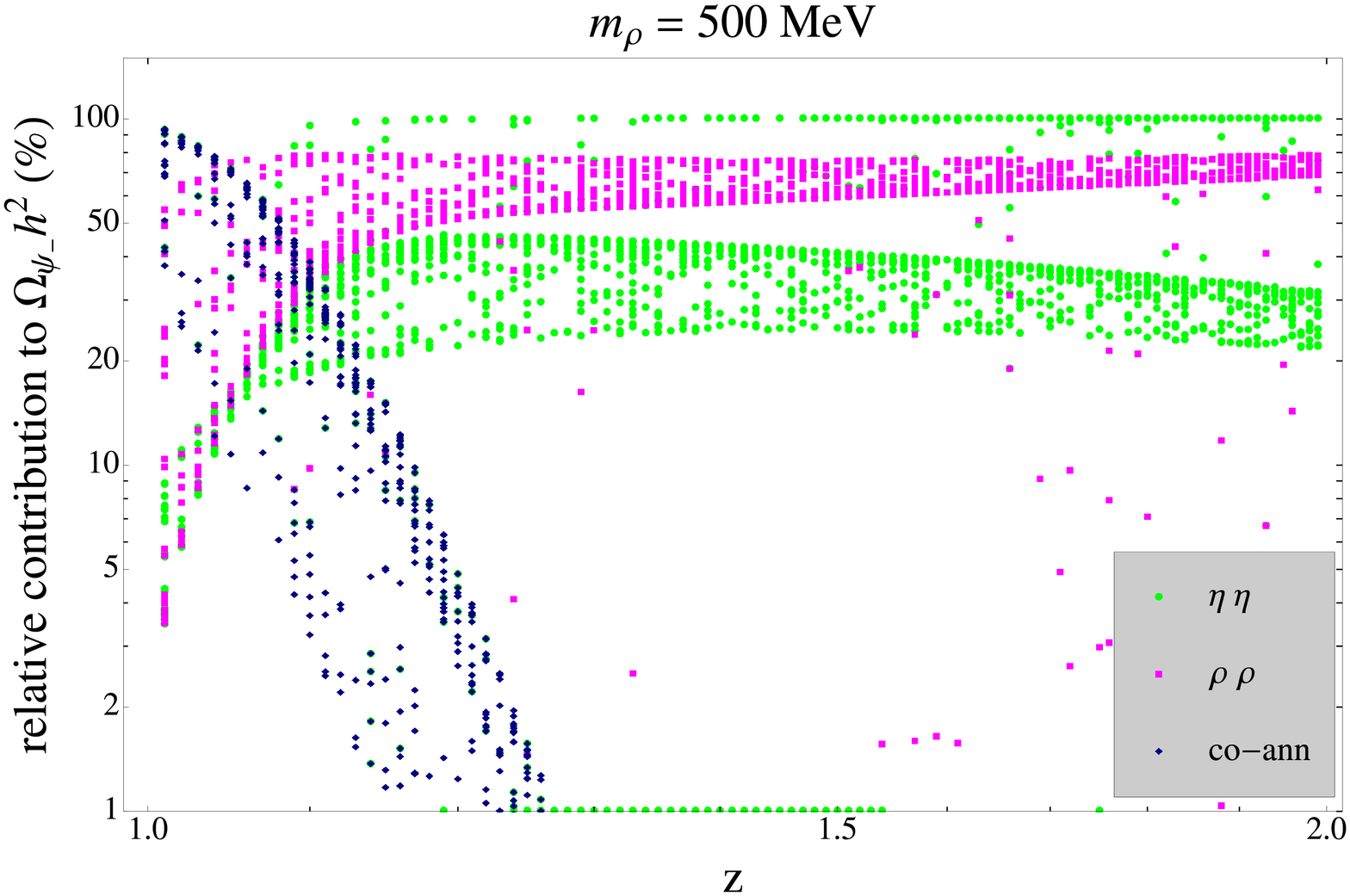}\\
\includegraphics[width=0.7\textwidth]{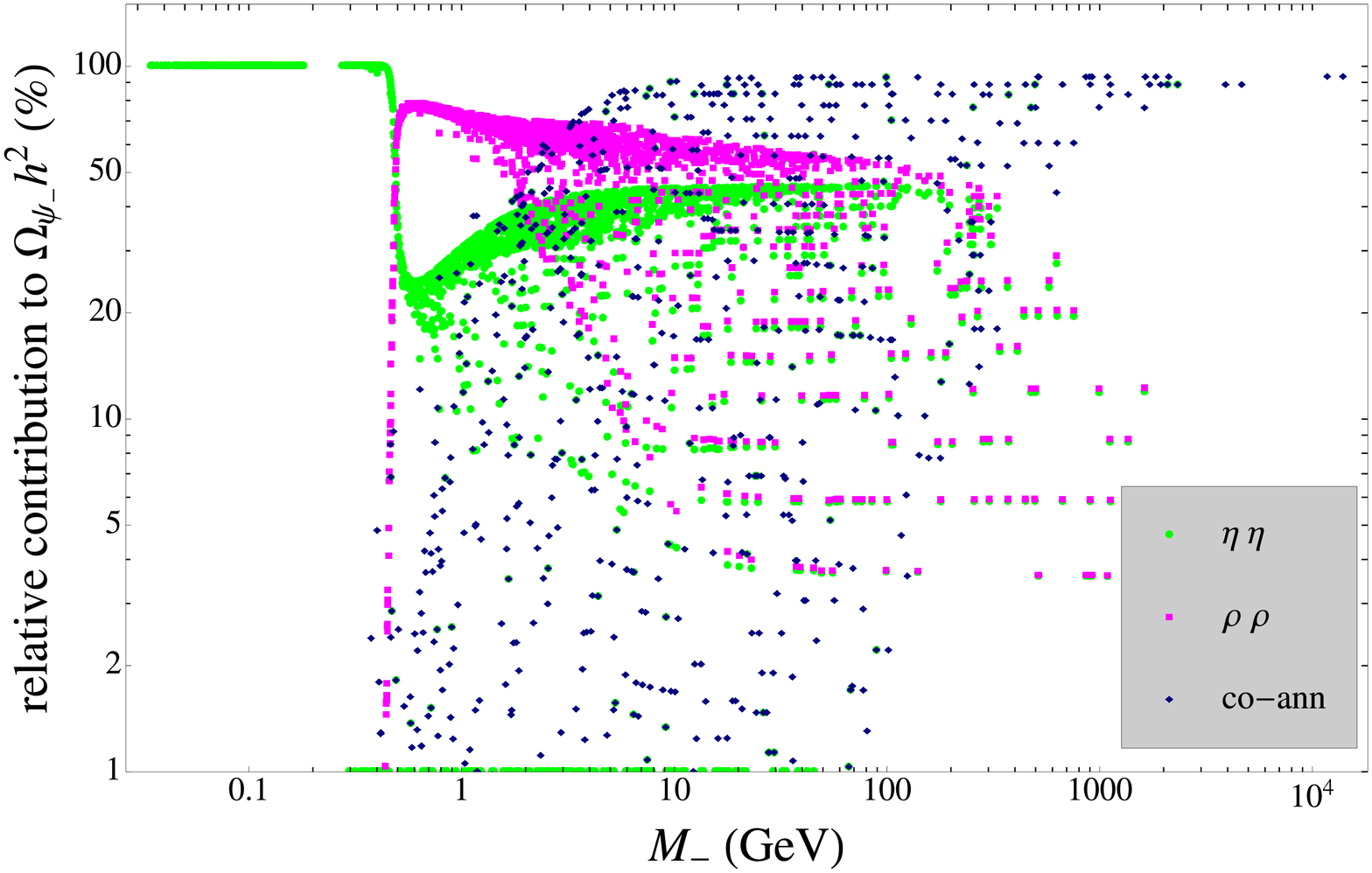}
\caption{Relative contribution of each annihilation channel to the dark matter relic density versus the degeneracy parameter $z$ (upper panel) and the dark matter mass  
(lower panel) for $m_{\rho}=500$ MeV. Only the the dark sector contributes to the relic density.}
\label{figure:rel1Dirac}
\end{center}
\end{figure}

\begin{figure}[t!]
\begin{center}
\includegraphics[width=0.7\textwidth]{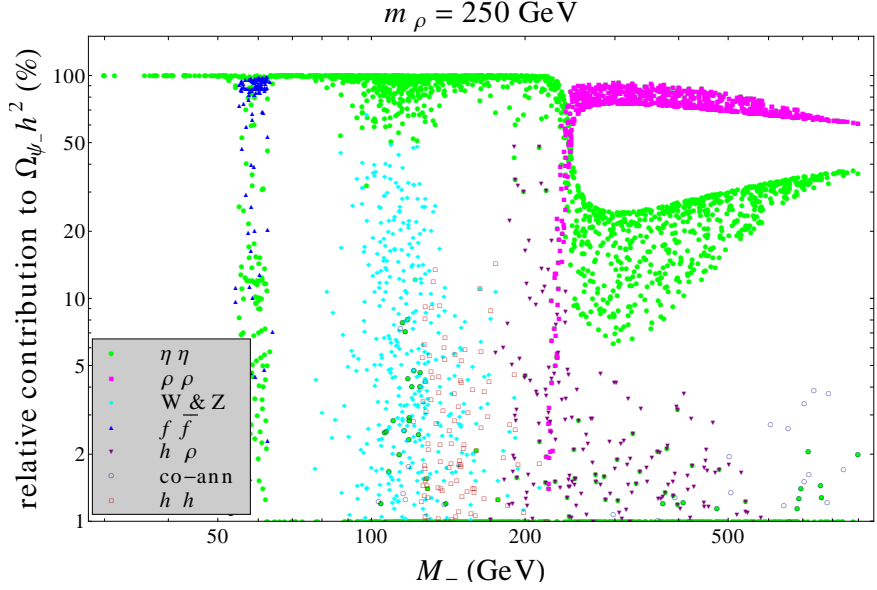}
\caption{Relative contribution of each annihilation channel to the dark matter relic density versus the dark matter mass for $m_{\rho}=250$ GeV.}
\label{figure:rel2Dirac}
\end{center}
\end{figure}

The dark matter relic abundance is obtained by solving the Boltzmann equation 
\begin{equation}
\frac{dn}{dt}+3 H n = -\langle\sigma_\text{eff} v \rangle \left(n^2 - (n^\text{eq})^2\right) \,
\label{boltzeq}
\end{equation}
where $n=n_++n_-$, with $n_\pm$ being the number densities of the (co)annihilating species $\psi_\pm$. The equilibrium densities are 
\begin{equation}
n_{\pm}^{\text{eq}} = \frac{M_\pm^2 T}{\pi^2} K_2\left(\frac{M_\pm}{T}\right)\,,
\end{equation}
where $K_{n}(x)$ is the modified Bessel function of the second kind.
The effective thermal cross-section is given by \cite{Gondolo:1990dk}
\begin{eqnarray}
\langle \sigma_\text{eff}  v \rangle  &=& \sum_{i,j=\pm}\langle\sigma^{ij} v \rangle \frac{n_i^{\text{eq}}}{n^\text{eq}}\frac{n_j^{\text{eq}}}{n^\text{eq}},
\label{sigmaeff}
\end{eqnarray}
with
\begin{eqnarray}
\langle \sigma^{ij}  v \rangle = \frac{\int^\infty_{(M_i+M_j)^2}  \frac{ds}{\sqrt{s}} K_1\left(\frac{\sqrt{s}}{T}\right) (s-(M_i+M_j)^2)(s-(M_i-M_j)^2) \sigma(ij \to all) }{ 8 T M_i^2 M_j^2 K_2(\frac{M_i}{T}) K_2(\frac{M_j}{T}) } \,.
\label{thermalsigma}
\end{eqnarray}

In order to study the dark matter production mechanism in this model, we perform  scans over the five dimensional parameter space. More specifically, $m_{\rho}$ and $v_\phi$ vary in the range 200 MeV and 1 TeV while the dark matter mass, between 1 GeV and 1 TeV.  Besides, $|\tan\theta|$ varies between zero and the upper limit set by the invisible decay width of the Higgs, given in Eq.~(\ref{thetabound}), and $f$ is taken between $10^{-2}$ and $4\pi$.  Lastly, for every point we require perturbativity, by requiring that the quartic couplings determined by Eq.~(\ref{kappaeq}) are smaller than 4$\pi$. For each of the points we then solve numerically the Boltzmann equation Eq.~(\ref{boltzeq}) by using micrOMEGAs 3.1~\cite{Belanger:2013oya}, working under an implementation of our model made with FeynRules~\cite{Christensen:2008py}, and select only those points for which the computed relic density is in agreement within 3$\sigma$ with the observed value $\Omega_{\text{DM}} h^2=0.1199\pm 0.0027$.

The results of two different scans, with $m_{\rho}=500$ MeV  and 250 GeV, are reported in Figs.~\ref{figure:rel1Dirac} and \ref{figure:rel2Dirac} respectively, where we show the relative contribution to the relic density of each annihilation channel. In the former,  the dominant annihilation channel is always into a pair of $\rho$ scalars, when this is kinematically open, $i.e.$ for $r=m_{\rho}/M_{-}<1$. Conversely, the annihilation into Goldstone bosons dominates in the region $r>1$. As shown in the upper panel, the coannihilation  process $\psi_{-}\psi_{+}\to \eta\rho$ is relevant only in the limit $z=M_{+}/M_{-}\sim 1$. For this region of the parameter space the annihilation of dark matter into the SM sector is always subdominant because either these channels are kinematically closed or the coupling to SM particles is very small. We show in Fig.~\ref{figure:rel2Dirac} the corresponding plot for the case $m_{\rho}=250$ GeV. In this case, due to the large scalar mass, annihilation channels into SM fermions and gauge bosons  might not be neglected. In general, channels with SM particles in the final state contribute significantly to the dark matter relic density only under two circumstances: when the mixing angle $\theta$ is non-negligible, or when the dark matter annihilation proceeds via resonant s-channel exchange of $CP$-even scalars, that is either $\rho$ or $h$. This is manifest in Fig.~\ref{figure:rel2Dirac} where the Higgs resonance takes place at $M_{-}= m_{h}/2\approx 63$ GeV while the one corresponding to $\rho$, at $M_{-}=m_{\rho}/2\approx 125$ GeV.

In the following we will focus in the limit $\theta\ll 1$, which is motivated by the measurements of the invisible Higgs decay width.  
In this regime, it is enough to consider the (co)annihilation channels with only $\eta$ or $\rho$ in the final state. In the first column of table \ref{table:annihilationdiagrams}, we show these channels and their corresponding Feynman diagrams. Some comments are in order:

\begin{itemize}
\item Not every channel is always kinematically allowed. In particular $\psi_-\psi_- \to \rho\rho$,  $\psi_+\psi_+ \to \rho\rho$ and  $\psi_-\psi_+ \to \eta\rho$ are only open if $m_\rho<M_- $, $m_\rho <M_+ $ and $m_\rho<(M_-+M_+)$ respectively, or equivalently if $r<1, r<z $ and $r<1+z$. Consequently,  a threshold effect associated to the opening of the channel $\psi_-\psi_- \to \rho\rho$ always takes place when $r \sim 1$. Furthermore, if coannihilations are relevant -- that is if $z \sim 1$--  a threshold effect appears also when $r\sim 2$ and $r \sim 1$, because of the opening of the channels $\psi_+\psi_- \to \eta\rho$ and $\psi_+\psi_+ \to \rho\rho$ respectively.  
\item $\psi_\pm \psi_\pm \to \eta\eta$ is always open and exhibits a resonant behavior when the energy of the initial state approaches the $\rho$ mass. This effect is more dramatic when $m_\rho>2 M_-$ --or equivalently when $r>2$-- because in this case the integration region of Eq.~(\ref{thermalsigma}) contains the resonance. However, this effect might also be present when $r>1$ because a part of the resonance peak might still be within the integration region. 
\end{itemize}
As a result, when  $r \gtrsim 0.8$ a full integration of the Boltzmann equations is needed without any approximation. In the Appendix, we show the exact formulas for the cross-sections which should be used in this case. Conversely, when $r \lesssim 0.8$, both resonance and threshold effects can be safely neglected and a reliable estimate of the cross-section can be derived from the so-called instantaneous freeze-out approximation by expanding in partial waves.  In table \ref{table:annihilationdiagrams} we report such expansion, that is,  we show the cross sections for each process to leading order in the relative velocity $v$ of the particles of the initial state, assuming a vanishing mixing angle $\theta$. The instantaneous freeze-out approximation reads \cite{Griest:1990kh} 
\begin{eqnarray}
	\Omega_{\text{DM}} h^{2} & \simeq & \frac{1.07\times 10^{9}\,\text{GeV}^{-1}}{J(x_{f})\, g_{*}(x_{f})^{1/2}\,m_{\text{Pl}}}\,,\label{omegah2}
\label{Omegah2}
\end{eqnarray}
where $x_{f} =  T_{f}/M_-$ and
\begin{eqnarray}
	J(x_{f}) &=& \int_{x_{f}}^{\infty}\,\frac{\langle\sigma_\text{eff} v\rangle}{x^{2}}\,\text{d}x\,, \hspace{20pt} x_{f} = \ln\frac{0.038\,c\,(c+2)\,g_{\text{eff}}\,m_{\text{Pl}}\,M_-\,\langle\sigma_\text{eff} v\rangle}{\left(g_{*}(x_{f})\,x_{f}\right)^{1/2}}\,.
\label{Jfxf}
\end{eqnarray}
Here $g_{*}(x_{f})$ is the number of relativistic degrees of freedom at the freeze-out temperature, $c$ is a constant of order one and 
$g_{\text{eff}}=2(1+n_{+}^{eq}/n_{-}^{eq})$. 
Typically, for a WIMP dark matter $x_{f}\approx 20-30$. 
In order to calculate the effective thermal cross-section $\langle\sigma_\text{eff} v\rangle$ in the non-relativistic limit, we use Eq.~(\ref{sigmaeff}) and 
\begin{equation}
\langle \sigma^{\pm\pm} v\rangle = a + 6\left(b-\frac{a}{4}\right)\frac{1}{x},\hspace{40pt} \text{if} \hspace{20pt}\sigma^{\pm\pm} v = a+b v^2 \,,
\end{equation}
which can be obtained from the expressions of table \ref{table:annihilationdiagrams}. 

As can be seen from table~\ref{table:annihilationdiagrams}, all the annihilation channels are p-wave suppressed. In contrast, the coannihilation channels  proceed via s-waves and are the dominant annihilation process in the early Universe if $z\sim 1$.  This can be understood from $CP$ conservation. To this end, we introduce the following notation: $L$ and $L_f$ are the orbital angular momenta of the initial and final states, respectively; $S$ and $S_f$ are the total spins of the initial and final states, respectively, and $J$ is the total angular momentum. Then, for the annihilation processes $\psi_-\psi_- \to \rho\rho$ and $\psi_-\psi_-\to \eta\eta$, the $CP$ eigenvalues of the initial and final states are $(-1)^{L+1}$ and $(-1)^{L_f}$, respectively. Thus $CP$-conservation implies that $|L_f - L|$ must be an odd number. In addition, since $\rho$ and $\eta$ are scalars, then $J=L_f$. If the s-wave were allowed, that is if $L=0$ or $J=S$, then we could only have $S=1$ and $L=0$, which is impossible for a pair of Majorana fermions due to the Pauli exclusion principle. The only possibility is therefore $L \geq 1$. On the other hand, for the coannihilation process $\psi_-\psi_+ \to \eta\rho$, the $CP$ eigenvalues of the initial and final states are $(-1)^{L}$ and $(-1)^{L_f+1}$. We again have $J=L_f$ and therefore $|J-L|$ must be an odd number. Consequently, the s-wave is allowed as long as $J=S=1$. Finally, we remark that the process $\psi_- \psi_- \to \rho \eta$ does not exist, despite $CP$ is conserved for some values of $L$. In fact, in  this process the initial state is  $C$-even whereas the final state is $C$-odd, hence it is forbidden by $C$ conservation. \\
\begin{table}[h]
	\centering
		\begin{tabular}{|c|c|} \hline
	{\bf Process} & {\bf Cross Section} \\\hline
Annihilation $\psi_-\psi_-\to \rho\rho$
& 
\multirow{2}{*}{ 
$\sigma v = \frac{f^4v^2}{16\pi M_-^2} G_{\rho\rho} (r,z)$
}\\
\multirow{6}{*}{
{
\unitlength=1.0 pt
\SetScale{1.0}
\SetWidth{0.7}      
\scriptsize    
\begin{picture}(96,38)(0,0)
\Line(48.0,35.0)(24.0,35.0) 
\Text(24.0,35.0)[r]{$\psi_-$}
\Line(48.0,35.0)(48.0,11.0) 
\Text(49.0,24.0)[l]{$\psi_-$}
\Text(49.0,27.0)[l]{$~$}
\DashLine(48.0,35.0)(72.0,35.0){3.0}
\Text(72.0,35.0)[l]{$\rho$}
\Line(48.0,11.0)(24.0,11.0) 
\Text(24.0,11.0)[r]{$\psi_-$}
\DashLine(48.0,11.0)(72.0,11.0){3.0}
\Text(72.0,11.0)[l]{$\rho$}
\end{picture}  
\begin{picture}(96,38)(0,0)
\Line(36.0,23.0)(12.0,35.0) 
\Text(12.0,35.0)[r]{$\psi_-$}
\Line(36.0,23.0)(12.0,11.0) 
\Text(12.0,11.0)[r]{$\psi_-$}
\DashLine(36.0,23.0)(60.0,23.0){3.0}
\Text(49.0,24.0)[b]{$\rho$}
\DashLine(60.0,23.0)(84.0,35.0){3.0}
\Text(84.0,35.0)[l]{$\rho$}
\DashLine(60.0,23.0)(84.0,11.0){3.0}
\Text(84.0,11.0)[l]{$\rho$}
\end{picture} 
}
} 
&\\ 
& $G_{\rho\rho} (r,z)\equiv \sqrt{1-r^2} (27 r^{12}+24 r^{10} z-240 r^{10}$\\
&
$+8 r^8 z^2-268 r^8 z+908 r^8-96 r^6 z^2+1152 r^6 z $\\
&
$-1920 r^6+420 r^4 z^2-2424 r^4 z+2436 r^4 $\\
&
$-800 r^2 z^2+2560 r^2 z-1760 r^2+576 z^2$\\
&
$-1152 z+576)/(6 (r^2-4)^2(r^2-2)^4 (z-1)^2)$ 
\\\hline
Annihilation $\psi_-\psi_-\to \eta\eta$
& 
\multirow{2}{*}{ 
$\sigma v = \frac{f^4v^2}{16\pi M_-^2} G_{\eta\eta} (r,z)$
} \\
\multirow{7}{*}{
{
\unitlength=1.0 pt
\SetScale{1.0}
\SetWidth{0.7}      
\scriptsize    
\begin{picture}(96,38)(0,0)
\Line(48.0,35.0)(24.0,35.0) 
\Text(24.0,35.0)[r]{$\psi_-$}
\Line(48.0,35.0)(48.0,11.0) 
\Text(49.0,24.0)[l]{$\psi_+$}
\DashLine(48.0,35.0)(72.0,35.0){3.0}
\Text(72.0,35.0)[l]{$\eta$}
\Line(48.0,11.0)(24.0,11.0) 
\Text(24.0,11.0)[r]{$\psi_-$}
\DashLine(48.0,11.0)(72.0,11.0){3.0}
\Text(72.0,11.0)[l]{$\eta$}
\end{picture}  
\begin{picture}(96,38)(0,0)
\Line(36.0,23.0)(12.0,35.0) 
\Text(12.0,35.0)[r]{$\psi_-$}
\Line(36.0,23.0)(12.0,11.0) 
\Text(12.0,11.0)[r]{$\psi_-$}
\DashLine(36.0,23.0)(60.0,23.0){3.0}
\Text(49.0,24.0)[b]{$\rho$}
\DashLine(60.0,23.0)(84.0,35.0){3.0}
\Text(84.0,35.0)[l]{$\eta$}
\DashLine(60.0,23.0)(84.0,11.0){3.0}
\Text(84.0,11.0)[l]{$\eta$}
\end{picture} 
}
} 
&\\
& $G_{\eta\eta} (r,z) \equiv 2 (3 r^4 z^4+2r^4 z^3+5 r^4 z^2+2 r^4$\\ 
&
$+12 r^2 z^6+4 r^2 z^5+8 r^2 z^4-8 r^2 z^3-12 r^2 z^2$\\
&
$+4 r^2 z-8 r^2+12 z^8-16 z^5-8 z^4+16 z^2$ \\
&
$-16 z+12)/(3 (r^2-4)^2 (z-1)^2(z^2+1)^4)$
\\\hline
Annihilation $\psi_+\psi_+\to \rho\rho$
& \\
\multirow{5}{*}{
{
\unitlength=1.0 pt
\SetScale{1.0}
\SetWidth{0.7}      
\scriptsize    
\begin{picture}(96,38)(0,0)
\Line(48.0,35.0)(24.0,35.0) 
\Text(24.0,35.0)[r]{$\psi_+$}
\Line(48.0,35.0)(48.0,11.0) 
\Text(49.0,24.0)[l]{$\psi_+$}
\Text(49.0,27.0)[l]{$~$}
\DashLine(48.0,35.0)(72.0,35.0){3.0}
\Text(72.0,35.0)[l]{$\rho$}
\Line(48.0,11.0)(24.0,11.0) 
\Text(24.0,11.0)[r]{$\psi_+$}
\DashLine(48.0,11.0)(72.0,11.0){3.0}
\Text(72.0,11.0)[l]{$\rho$}
\end{picture}  
\begin{picture}(96,38)(0,0)
\Line(36.0,23.0)(12.0,35.0) 
\Text(12.0,35.0)[r]{$\psi_+$}
\Line(36.0,23.0)(12.0,11.0) 
\Text(12.0,11.0)[r]{$\psi_+$}
\DashLine(36.0,23.0)(60.0,23.0){3.0}
\Text(49.0,24.0)[b]{$\rho$}
\DashLine(60.0,23.0)(84.0,35.0){3.0}
\Text(84.0,35.0)[l]{$\rho$}
\DashLine(60.0,23.0)(84.0,11.0){3.0}
\Text(84.0,11.0)[l]{$\rho$}
\end{picture} 
}
} 
&\\ 
&$\sigma v = \frac{f^4v^2}{16\pi M_-^2z^2} G_{\rho\rho} (\frac{r}{z},\frac{1}{z})$\\ 
&\\
&
\\\hline
Annihilation $\psi_+\psi_+\to \eta\eta$
&\\
\multirow{5}{*}{
{
\unitlength=1.0 pt
\SetScale{1.0}
\SetWidth{0.7}      
\scriptsize    
\begin{picture}(96,38)(0,0)
\Line(48.0,35.0)(24.0,35.0) 
\Text(24.0,35.0)[r]{$\psi_+$}
\Line(48.0,35.0)(48.0,11.0) 
\Text(49.0,24.0)[l]{$\psi_-$}
\DashLine(48.0,35.0)(72.0,35.0){3.0}
\Text(72.0,35.0)[l]{$\eta$}
\Line(48.0,11.0)(24.0,11.0) 
\Text(24.0,11.0)[r]{$\psi_+$}
\DashLine(48.0,11.0)(72.0,11.0){3.0}
\Text(72.0,11.0)[l]{$\eta$}
\end{picture}  
\begin{picture}(96,38)(0,0)
\Line(36.0,23.0)(12.0,35.0) 
\Text(12.0,35.0)[r]{$\psi_+$}
\Line(36.0,23.0)(12.0,11.0) 
\Text(12.0,11.0)[r]{$\psi_+$}
\DashLine(36.0,23.0)(60.0,23.0){3.0}
\Text(49.0,24.0)[b]{$\rho$}
\DashLine(60.0,23.0)(84.0,35.0){3.0}
\Text(84.0,35.0)[l]{$\eta$}
\DashLine(60.0,23.0)(84.0,11.0){3.0}
\Text(84.0,11.0)[l]{$\eta$}
\end{picture} 
}
} 
&\\
&$\sigma v = \frac{f^4v^2}{16\pi M_-^2z^2} G_{\eta\eta} (\frac{r}{z},\frac{1}{z})$\\
&\\
& 
\\\hline
Coannihilation $\psi_-\psi_+ \to \rho\eta$
& 
\multirow{3}{*}{ 
$\sigma v = \frac{f^4}{16\pi M_-^2} G_{\rho\eta} (r,z)$ 
}\\
\multirow{4}{*}{ 
{
\unitlength=1.0 pt
\SetScale{1.0}
\SetWidth{0.7}      
\scriptsize    
\begin{picture}(96,38)(0,0)
\Line(48.0,35.0)(24.0,35.0) 
\Text(24.0,35.0)[r]{$\psi_-$}
\Line(48.0,35.0)(48.0,11.0) 
\Text(49.0,24.0)[l]{$\psi_+$}
\DashLine(48.0,35.0)(72.0,35.0){3.0}
\Text(72.0,35.0)[l]{$\eta$}
\Line(48.0,11.0)(24.0,11.0) 
\Text(24.0,11.0)[r]{$\psi_+$}
\DashLine(48.0,11.0)(72.0,11.0){3.0}
\Text(72.0,11.0)[l]{$\rho$}
\end{picture}  
\begin{picture}(96,38)(0,0)
\Line(36.0,23.0)(12.0,35.0) 
\Text(12.0,35.0)[r]{$\psi_-$}
\Line(36.0,23.0)(12.0,11.0) 
\Text(12.0,11.0)[r]{$\psi_+$}
\DashLine(36.0,23.0)(60.0,23.0){3.0}
\Text(49.0,24.0)[b]{$\eta$}
\DashLine(60.0,23.0)(84.0,35.0){3.0}
\Text(84.0,35.0)[l]{$\eta$}
\DashLine(60.0,23.0)(84.0,11.0){3.0}
\Text(84.0,11.0)[l]{$\rho$}
\end{picture} 
}
}
&\\
&\\
& $G_{\rho\eta} (r,z)\equiv((z+1)^2-r^2)^5/(4 (r^2 z-z^2$\\
\multirow{3}{*}{
{
\unitlength=1.0 pt
\SetScale{1.0}
\SetWidth{0.7}      
\scriptsize    
\begin{picture}(96,38)(0,0)
\Line(48.0,35.0)(24.0,35.0) 
\Text(24.0,35.0)[r]{$\psi_-$}
\Line(48.0,35.0)(48.0,11.0) 
\Text(49.0,24.0)[l]{$\psi_-$}
\DashLine(48.0,35.0)(72.0,35.0){3.0}
\Text(72.0,35.0)[l]{$\rho$}
\Line(48.0,11.0)(24.0,11.0) 
\Text(24.0,11.0)[r]{$\psi_+$}
\DashLine(48.0,11.0)(72.0,11.0){3.0}
\Text(72.0,11.0)[l]{$\eta$}
\end{picture} 
}
}
&$-2 z-1)^2 (r^2-z^3-2 z^2-z)^2)$ \\
&\\
&\\\hline
	\end{tabular}
\caption{ (Co)annihilation channels of the dark matter particle and the corresponding cross-sections to leading order in the relative velocity $v$ and the mixing angle $\theta$.
The exact expressions are reported in the Appendix.}
\label{table:annihilationdiagrams}
\end{table}
\begin{figure}[t]
\begin{center}
\includegraphics[width=8cm]{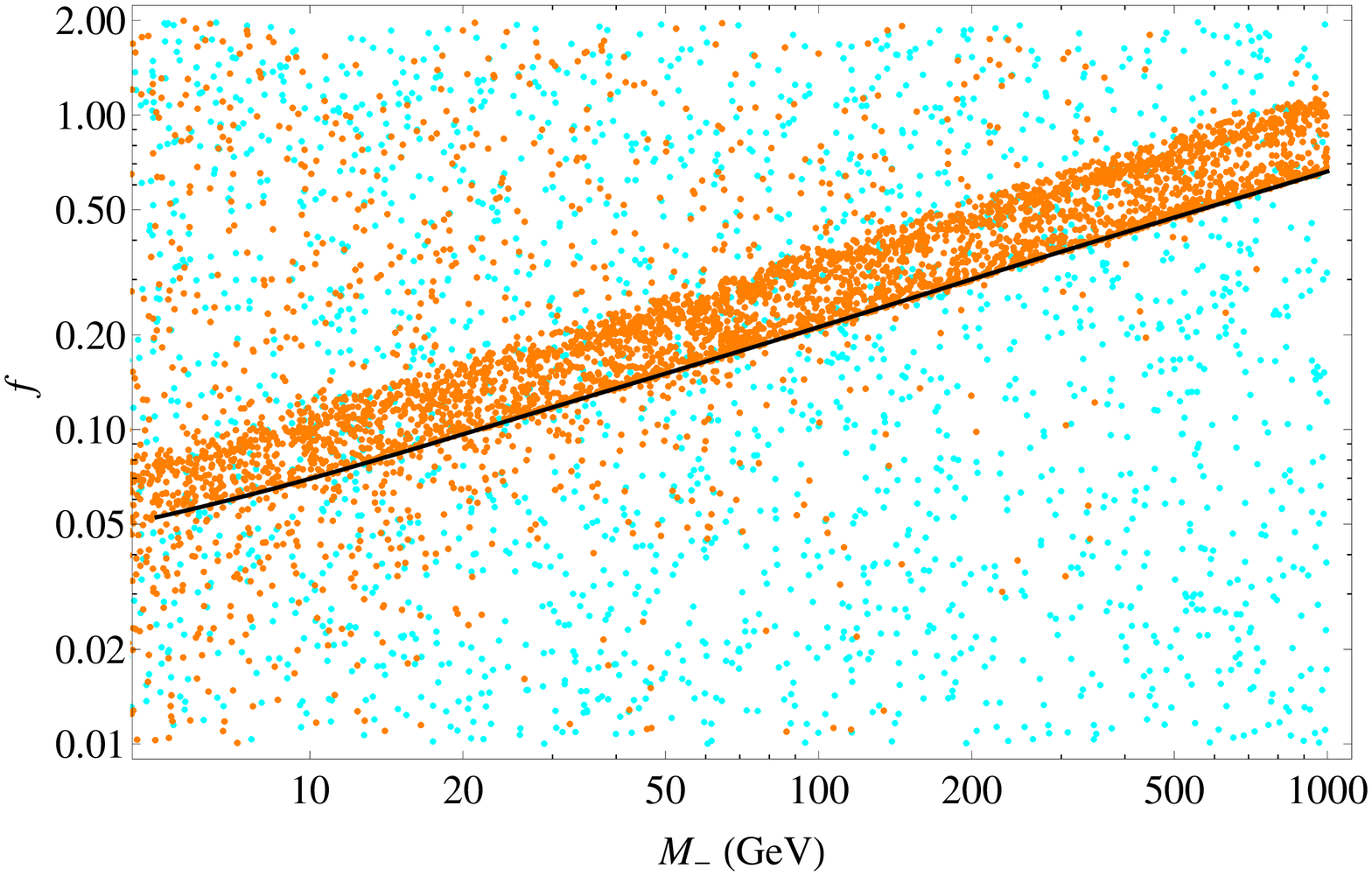}
\includegraphics[width=8cm]{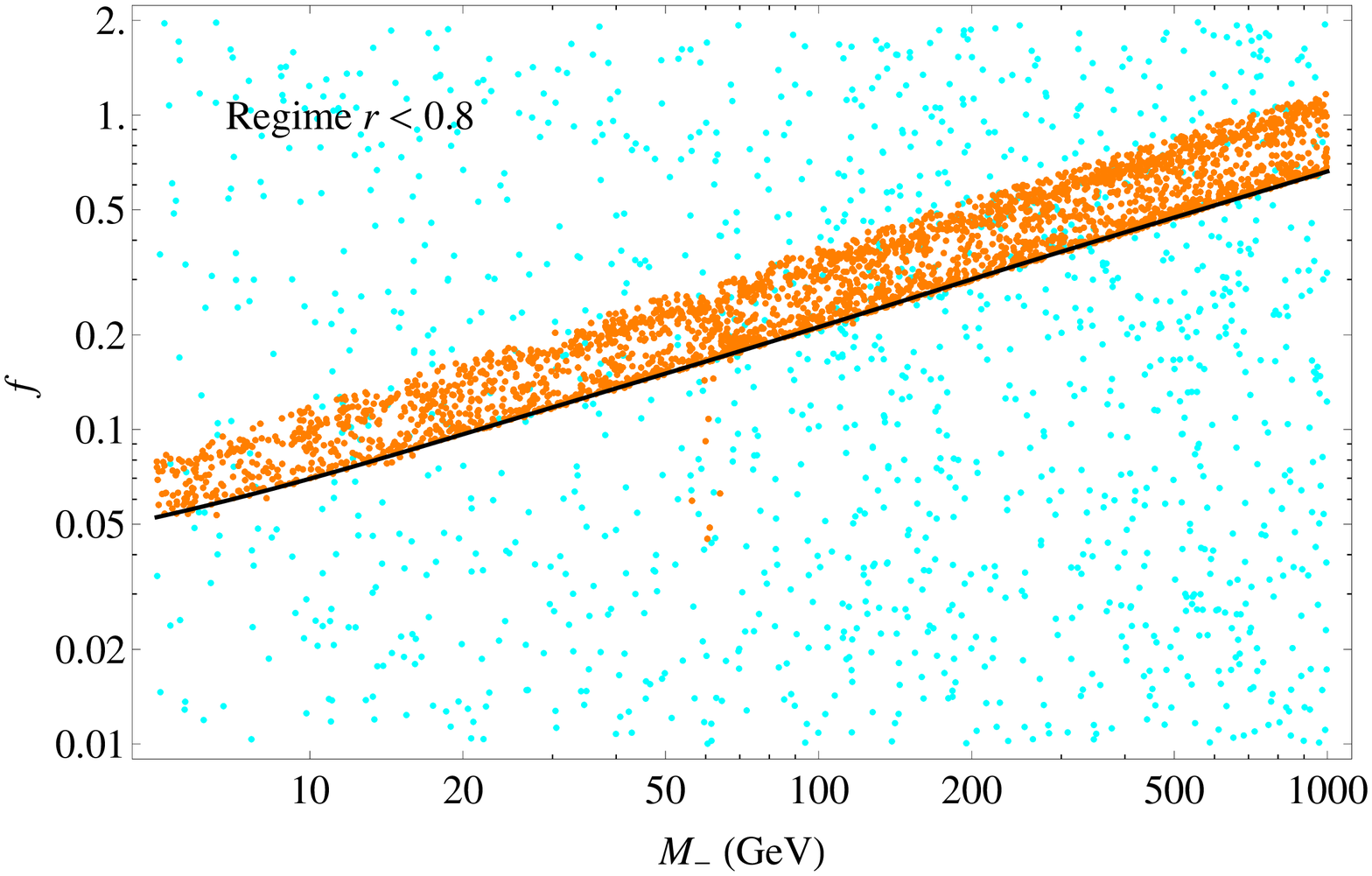}
\caption{Coupling constant $f$ versus dark matter mass for random points of the five dimensional parameter space (see the text for details). Only the orange points 
reproduce the observed relic density.
The black line in the right panel corresponds to the coannihilation limit given in Eq.~(\ref{UpperLimitf}).}
\label{figure:fvsM}
\end{center}
\end{figure}

For the points that reproduce the correct relic density we expect, in the regime where $r\lesssim 0.8$, a lower limit on the coupling $f$ as a function of the dark matter mass, corresponding to the points where $z\sim 1$, namely to the coannihilation limit. In this case, 
annihilations are p-wave suppressed while coannihilations are not. Consequently, in the former case larger values of $f$ are required in order to reproduce the same total annihilation cross section.
The lower limit can be analytically estimated using that $\langle\sigma_{\text{eff}} v\rangle \to f^{4}/(32\,\pi\,M_{-}^{2})$ when $z\to1$. Then, Eq.~(\ref{Omegah2}) simplifies to
\begin{eqnarray} 
	\Omega_{\text{DM}} h^{2}\Bigg|_{z\to 1} & \simeq & \frac{1.07\times 10^{11}\,\text{GeV}^{-1} x_f\,M_-^2}{
 \,f^{4}\, g_{*}(x_{f})^{1/2}\,m_{\text{Pl}}}\,.
\label{Omegah2zto1}
\end{eqnarray}
Furthermore, we can solve for $f$ as a function of $M_-$
\begin{eqnarray}
	f\Bigg|_{z\to 1} & \simeq &\left( \frac{1.07\times 10^{11}\,\text{GeV}^{-1} \,x_f}{
 \, g_{*}(x_{f})^{1/2}\,m_{\text{Pl}}\,\Omega_{\text{DM}} h^{2}} \right)^{1/4} M_-^{1/2} \,.
\label{UpperLimitf}
\end{eqnarray}
which corresponds to the lower bound on the coupling constant $f$.  We show in Fig.~\ref{figure:fvsM} as cyan points the values of the coupling constant $f$ versus the dark matter mass $M_-$ obtained from a scan over the five dimensional parameter space following the procedure described before; the orange points correspond to the subset of points that reproduce the observed relic density $\Omega_{\text{DM}} h^2=0.1199\pm 0.0027$. In the left panel we include all points, whereas in the right panel we show only those for which $r\lesssim 0.8$.  
From the right plot, it is apparent the correlation between the coupling and the dark matter mass, as well as the existence of a lower limit on the coupling. This lower limit is reasonably well reproduced by the analytic expression reported in Eq.~(\ref{UpperLimitf}), calculated for $x_f = 25$ and shown in the plot as a black line, except for
the orange points  around $M_{-}=m_{h}/2\simeq 63$ GeV, due to the Higgs resonance, where Eq.~(\ref{Omegah2zto1}) does not hold.
In contrast,  a correlation does not exist in the left plot, due to the presence of resonance and threshold effects.

\section{Constraints from Direct Detection Experiments}\label{DirectDetection}
\label{sec:direct_detection}

\begin{figure}[t]
%
%
%
%
\begin{center}
{
\unitlength=1.0 pt
\SetScale{1.0}
\SetWidth{0.7}      
\scriptsize    
{} \quad\allowbreak
\begin{picture}(96,38)(0,0)
\Text(65.0,45.0)[r]{$-\frac{if}{2}\sin\theta$}
\Text(65.0,0.0)[r]{$-\frac{im_q}{v_H}\cos\theta$}
\Line(9.0,35.0)(48.0,35.0) 
\Text(9.0,35.0)[r]{$\psi_-$}
\DashLine(48.0,35.0)(48.0,11.0){3}
\Text(49.0,24.0)[l]{$h$}
\Line(48.0,35.0)(87.0,35.0) 
\Text(87.0,35.0)[l]{$\psi_-$}
\ArrowLine(9.0,11.0)(48.0,11.0)
\Text(7.0,11.0)[r]{$q$}
\ArrowLine(48.0,11.0)(87.0,11.0) 
\Text(88.0,11.0)[l]{$q$}
\end{picture} \ 
{} \quad\allowbreak
\begin{picture}(96,38)(0,0)
\Text(65.0,45.0)[r]{$\frac{if}{2}\cos\theta$}
\Text(65.0,0.0)[r]{$-\frac{im_q}{v_H}\sin\theta$}
\Line(9.0,35.0)(48.0,35.0) 
\Text(9.0,35.0)[r]{$\psi_-$}
\DashLine(48.0,35.0)(48.0,11.0){3}
\Text(49.0,24.0)[l]{$\rho$}
\Line(48.0,35.0)(87.0,35.0) 
\Text(87.0,35.0)[l]{$\psi_-$}
\ArrowLine(9.0,11.0)(48.0,11.0)
\Text(7.0,11.0)[r]{$q$}
\ArrowLine(48.0,11.0)(87.0,11.0) 
\Text(88.0,11.0)[l]{$q$}
\end{picture} \ 
}

\end{center}
\caption{Relevant Feynman diagrams for dark matter direct detection experiments. }
\label{figure:DDdiagrams}
\end{figure}
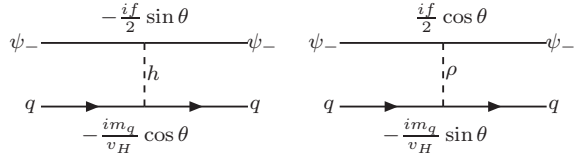

In Fig.~\ref{figure:DDdiagrams}, we show the diagrams that are relevant for dark matter direct detection experiments. Following \cite{Jungman:1995df}, we calculate the corresponding WIMP-nucleon scattering cross-section
\begin{eqnarray}
\sigma_{\psi_-N}& =& C^2\, \frac{ f^2\, m_N^4 \,M_-^2}{4\pi \,v_H^2\, (M_-+m_N)^2} \,\left(\frac{1}{m_h^2}-\frac{1}{m_\rho^2}\right)^2 \sin^22\theta\,,
\label{DirectDetectionEq}
\end{eqnarray}
where $m_N$ denotes the nucleon mass and $C\simeq 0.27$ \cite{Belanger:2013oya} is a constant that depends on the nucleon matrix element. As shown in Fig.~\ref{figure:DDdiagrams}, there is a relative sign between the Higgs and the $\rho$ particle amplitudes, which is responsible for the destructive interference term in Eq.~(\ref{DirectDetectionEq}). Note that the scattering cross section has a strong dependence with  $m_\rho$ (concretely with $m_\rho^{-4}$) when $m_\rho< m_h$, while it is independent of $m_\rho$ when $m_\rho>m_h$. These two limiting behaviors correspond to the regimes where the scattering is dominated by the $\rho$ scalar or by the Higgs boson, respectively.  Besides, the scattering cross section is suppressed when $m_\rho\simeq m_h$. 

\begin{figure}[t]
\begin{center}
\includegraphics[width=8cm]{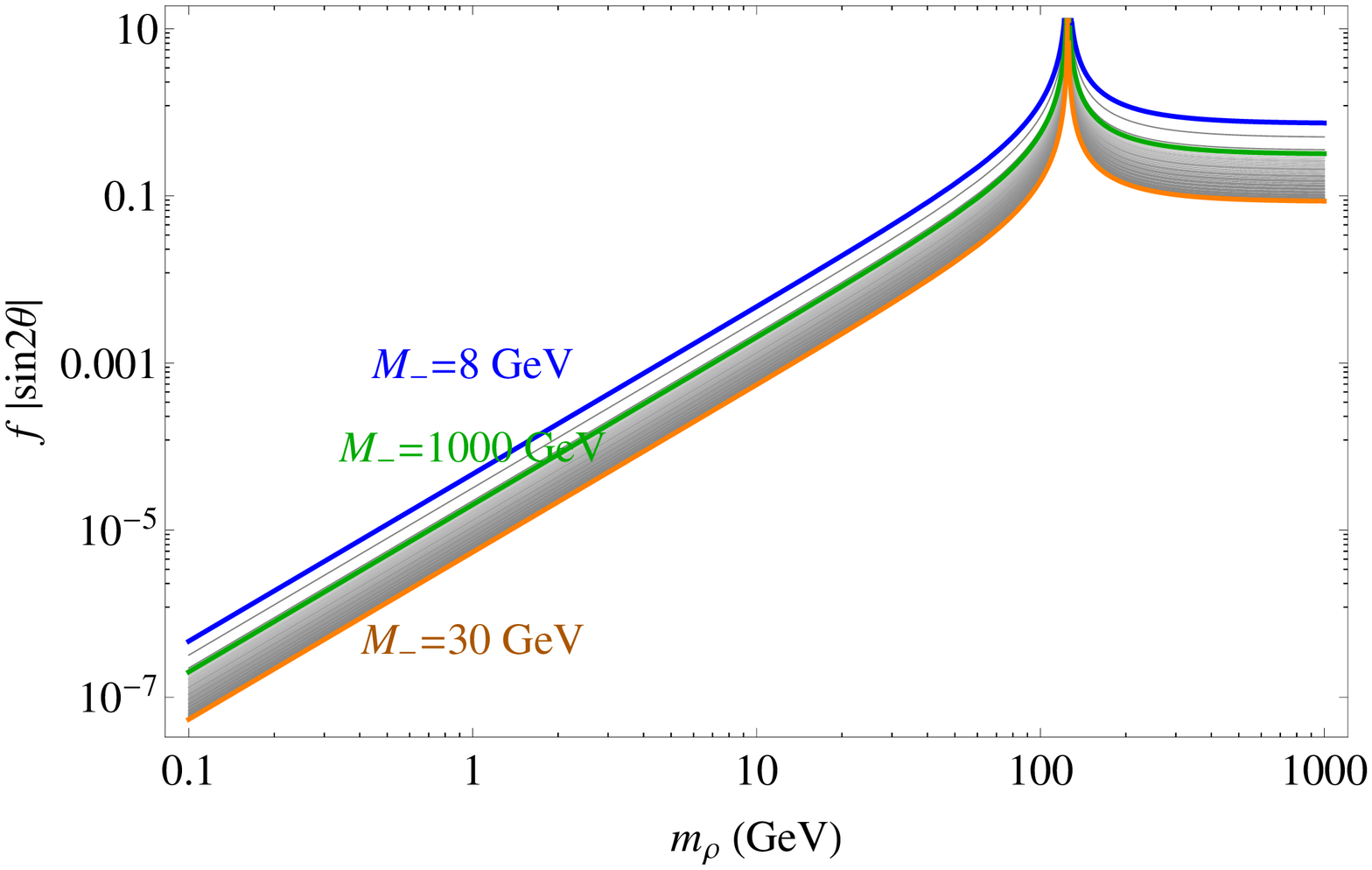} 
\includegraphics[width=8cm]{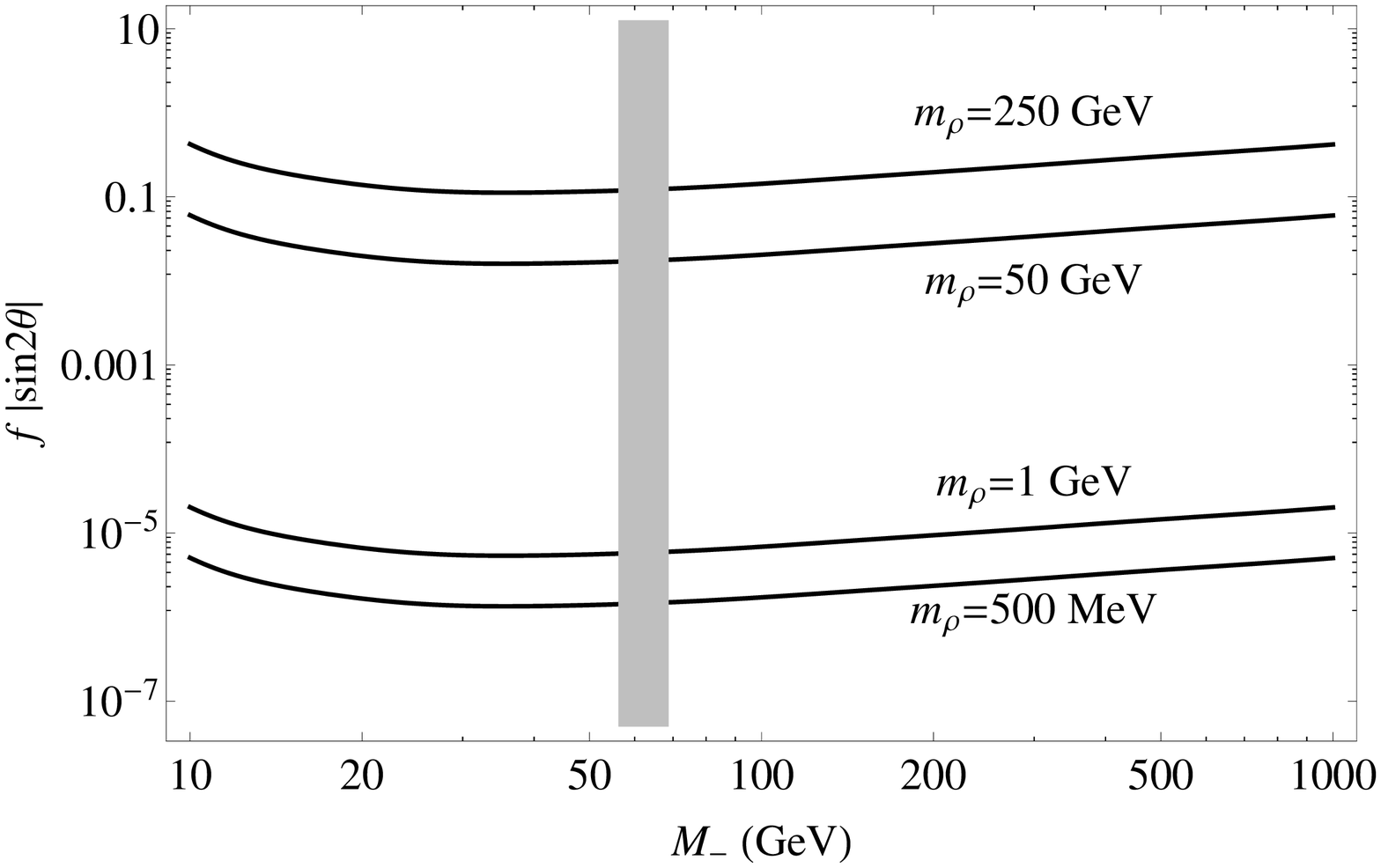}
\caption{Upper bound on $f|\sin2\theta|$ from LUX current limits on WIMP-nucleon cross-sections (see the text for details).  }
\label{figure:fsin2thetaBound}
\end{center}
\end{figure}

The limits on the scattering cross section of dark matter particles with protons from the  LUX experiment \cite{Akerib:2013tjd} translate into limits on the parameter space of our scenario. In the left panel of Fig.~\ref{figure:fsin2thetaBound} we show, as black lines, the bounds on $f|\sin2\theta|$ as a function of $m_\rho$ for various dark matter masses in between $8 ~\GeV$ and $1000 ~\GeV$; in blue, orange and green we show the bound for $M_-=8, 30$ and $1000~\GeV$ respectively. The limits are stronger for dark matter masses close to $30~\GeV$, as a result of the larger sensitivity of the LUX experiment to WIMP masses around this value. Also, the dependence of the cross section with $m_\rho$ described above is reflected in the bound on $f|\sin2\theta|$, as apparent from the plot.

\begin{figure}[t]
\begin{center}
\includegraphics[width=8cm]{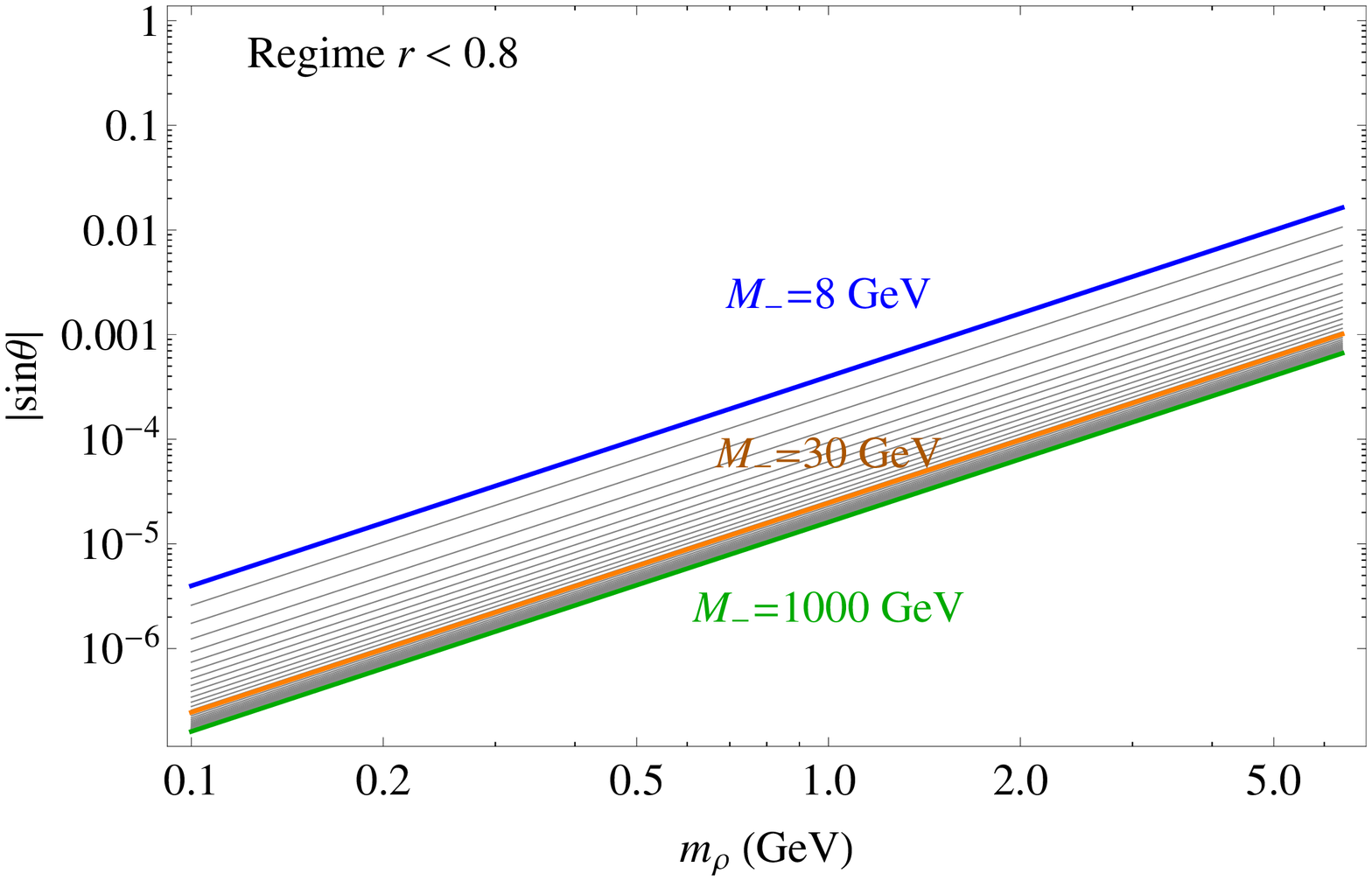} 
\includegraphics[width=8cm]{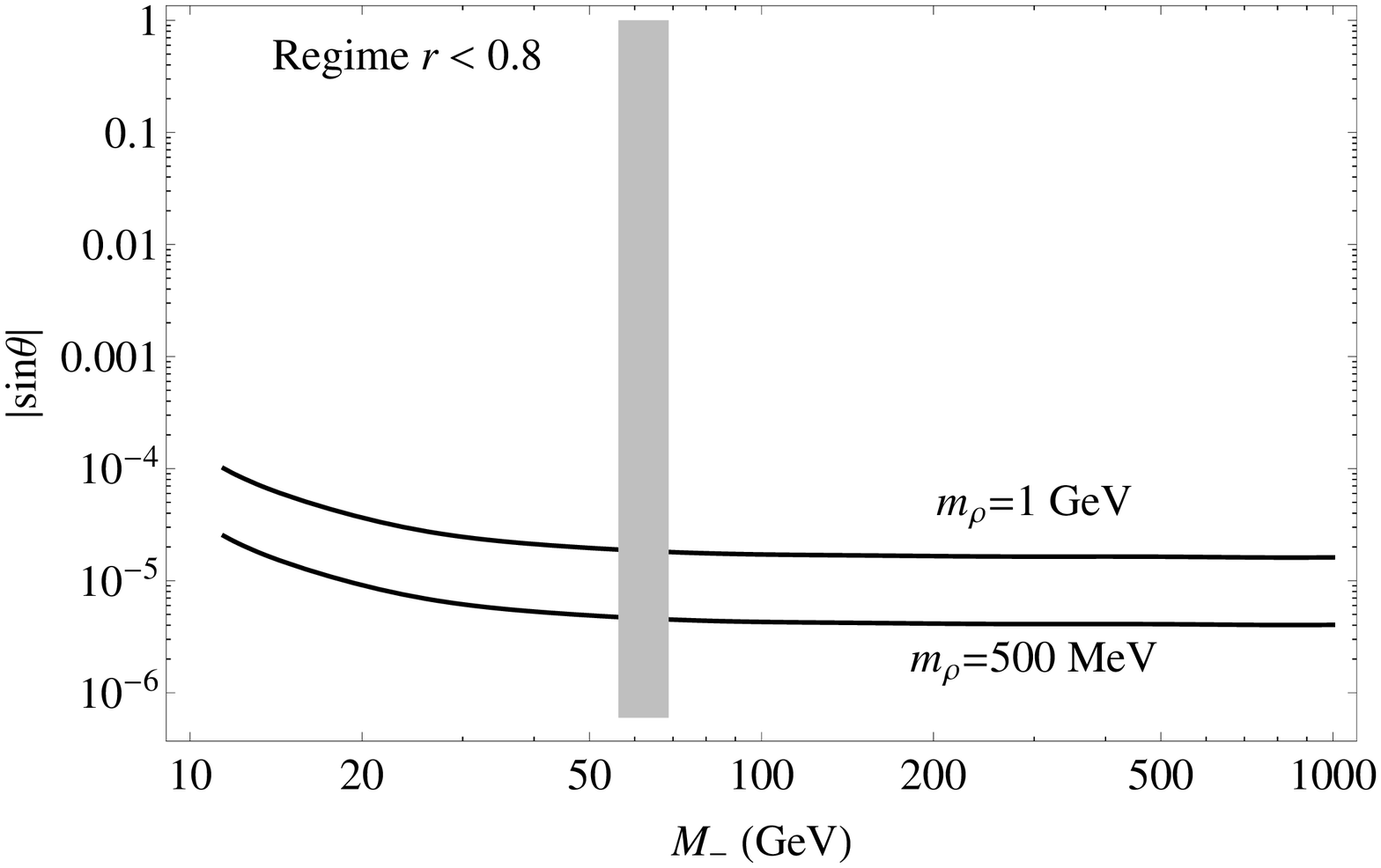}
\caption{Same as Fig.~\ref{figure:fsin2thetaBound}, but for $|\sin\theta|$ in the region $r\lesssim 0.8$ (see text for details).  }
\label{figure:sin2thetaBound}
\end{center}
\end{figure}

For small values of $m_\rho$ - namely smaller than about $6~\GeV$ - the dark matter masses probed by LUX satisfy $r \lesssim 0.8$. Hence, if the dark matter was produced thermally,  the lower limit on $f$ of Eq.~(\ref{UpperLimitf}) can be applied in order to get an upper bound on $|\sin\theta|$. This is shown in Fig.~\ref{figure:sin2thetaBound}. As before, the left panel shows, as black lines, the upper limits on $|\sin\theta|$ as a function of $m_\rho$ for various dark matter masses between $8 ~\GeV$ and $1000 ~\GeV$; the limits for the concrete masses $M_-=8, 30$ and $1000 ~\GeV$ are shown in blue, orange and green, respectively. Besides, in the right panel of Fig.~\ref{figure:sin2thetaBound}, we report the same limits as a function of the dark matter mass for fixed values of $m_\rho$. Notice that, above dark matter masses of about  $30 ~\GeV$, the LUX limits on $|\sin\theta|$ are almost independent of the dark matter mass. In fact, in the left panel, the green and the orange lines almost coincide. The region around the Higgs resonance in the right panel is not included in the analysis since Eq.~(\ref{Omegah2zto1}) does not apply. 


\section{Goldstone Bosons as Dark Radiation}
\label{sec:dark_radiation}

Since the Goldstone bosons are massless particles, they contribute to the radiation energy density of the Universe.
In particular, as pointed out in \cite{Weinberg:2013kea}, if they go out of equilibrium before the annihilation of the $e^{\pm}$ pairs, but after the decoupling of most of the SM fermions,
they might fake extra neutrino species in the measurements of the anisotropies in the
cosmic microwave background (CMB) \cite{Planck2013}. This effect  can be  quantified in terms of the effective number of neutrino types, $N_{eff}$, present before the era of 
recombination. 

Following \cite{Steigman}, we define $T^{0}_{\eta}$ and $T^{d}_{\eta}$  as the temperature of the Goldstone bosons today and at their decoupling from the thermal bath, respectively. A similar notation is understood for all the other particles.  In terms of these quantities, $N_{eff}$ is given by:
\begin{equation}
	N_{eff} \;=\; 3\,\left(1\,+
	\,\frac{\Delta N_{\eta}}{3}\,\left(\frac{T_{\eta}^{0}}{T_{\nu}^{0}}\right)^{4}\right)\,,\label{Neffdef}
\end{equation}
where $\Delta N_{\eta}= 4/7$ because  the Goldstone boson is the only scalar decoupled at $T>T^{d}_{\eta}$ and present before the recombination era.

In the present scenario we expect the massless scalars to be very weakly interacting with the SM particles, in fact even more than neutrinos.
Therefore, they should decouple at a temperature $T_{\eta}^{d}>T_{\nu}^{d}$, with $T_{\nu}^{d}\simeq2-3$ MeV \cite{Enqvist, Dolgov, Hannestad}, which in turn implies that
 the ratio $T_{\eta}/T_{\nu}$ today is the same as it was at $T_{\nu}^{d}$
because  neutrinos and Goldstone bosons have been decoupled from the thermal bath ever since the temperature dropped below $T_{\nu}^{d}$. 
Moreover, the temperature of the neutrinos and the Goldstone bosons are not the same at $T_{\nu}^{d}$ because  in between 
the Goldstone boson and neutrino decoupling epochs,
the thermal bath underwent a reheating process due to the annihilations of some of the fermions in the plasma. This effect can be quantified
by considering the conservation of the entropy per comoving volume during that period of time.
This implies that $g_{*}(T)\,T^{3}$ remained constant, where $g_{*}\left(T\right)$ stands for
the effective number of relativistic degrees of freedom.
As a consequence of all this, we have
\begin{equation}
	\left(\frac{T_{\eta}^{0}}{T_{\nu}^{0}}\right)^{3}\;=\; \left(\frac{T_{\eta}}{T_{\nu}}\right)_{T_{\nu}^{d}}^{3}\;=\;
	\frac{g_{*}\left(T_{\nu}^{d}\right)}{g_{*}\left(T_{\eta}^{d}\right)}\,,
\end{equation}
 with\footnote{This results is strictly valid under the assumption that the neutrinos decouple instantaneously and that at $T_{\nu}^{d}$
the $e^{\pm}$ pairs are essentially massless \cite{Mangano:2005cc}.} 
$g_{*}(T_{\nu}^{d})=43/4=10.75$. Then,
in terms of the Goldstone decoupling temperature the effective neutrino number results \cite{Steigman}
\begin{equation}
	N_{eff} \;=\;
	3\,\left(1\,+
	\,\frac{\Delta N_{\eta}}{3}\,\left(\frac{g_{*}\left(T_{\nu}^{d}\right)}{g_{*}\left(T_{\eta}^{d}\right)}\right)^{4/3}\right)\,.\label{Neff}
\end{equation}
In this work we assume that the Goldstone bosons 
decouple just before muon annihilation. As a result $g_{*}\left(T_{\eta}^{d}\right)=57/4$, which from Eq.~(\ref{Neff}) corresponds to an effective number 
of neutrinos  $N_{eff}-3=(4/7)(43/57)^{4/3}\simeq  0.39$  \cite{Weinberg:2013kea}.  
As pointed out before, the latter is consistent within $1\sigma$ with the recent experimental data \cite{Planck2013}, $N_{eff}=3.36\pm 0.34$.

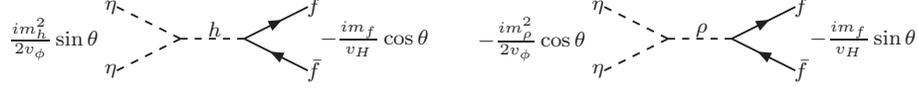
\begin{figure}[t]
\begin{center}
{
\unitlength=1.0 pt
\SetScale{1.0}
\SetWidth{0.7}      
\scriptsize    
{} \qquad\allowbreak
\begin{picture}(96,38)(0,0)
\DashLine(36.0,23.0)(12.0,35.0){3}
\Text(12.0,35.0)[r]{$\eta$}
\DashLine(36.0,23.0)(12.0,11.0){3}
\Text(12.0,11.0)[r]{$\eta$}
\DashLine(36.0,23.0)(60.0,23.0){3}
\Text(49.0,24.0)[b]{$h$}
\ArrowLine(60.0,23.0)(84.0,35.0) 
\Text(84.0,35.0)[l]{$f$}
\ArrowLine(84.0,11.0)(60.0,23.0) 
\Text(84.0,11.0)[l]{$\bar{f}$}
\Text(5.0,23.0)[r]{$\frac{im_h^2}{2v_\phi}\sin\theta$}
\Text(130.0,23.0)[r]{$-\frac{im_f }{v_H}\cos\theta$}

\end{picture} \ 
{} \qquad\allowbreak
\hspace{60pt}
\begin{picture}(96,38)(0,0)
\DashLine(36.0,23.0)(12.0,35.0){3}
\Text(12.0,35.0)[r]{$\eta$}
\DashLine(36.0,23.0)(12.0,11.0){3}
\Text(12.0,11.0)[r]{$\eta$}
\DashLine(36.0,23.0)(60.0,23.0){3}
\Text(49.0,24.0)[b]{$\rho$}
\ArrowLine(60.0,23.0)(84.0,35.0) 
\Text(84.0,35.0)[l]{$f$}
\ArrowLine(84.0,11.0)(60.0,23.0) 
\Text(84.0,11.0)[l]{$\bar{f}$}
\Text(5.0,23.0)[r]{$-\frac{im_\rho^2 }{2v_\phi}\cos\theta$}
\Text(130.0,23.0)[r]{$-\frac{im_f }{v_H}\sin\theta$}
\end{picture} \ 
}

\end{center}
\caption{Relevant Feynman diagrams for Goldstone boson annihilation into SM fermions. }
\label{figure:etadecdiagrams}
\end{figure}

In order to analyze carefully the conditions under  which the Goldstone bosons decouple from the thermal bath, we consider the  Boltzmann equation describing the evolution in the early Universe of the Goldstone boson number density, $n_{\eta}$. We assume for simplicity that the $\rho$ scalar and the dark matter are no longer present in the thermal bath at the decoupling of the Goldstone. If that is the case, the evolution of $n_{\eta}$ is described by  
\begin{equation}
	\frac{dn_{\eta}}{dt}\,+\,3\,H\,n_{\eta}\;=\;-\,\sum_{f}\,\langle \sigma v \rangle_{\eta\eta\to f \bar{f}} \,\left(\,n^{2}_{\eta}\,-\,(n_{\eta}^{eq})^{2}\,\right)\,,\label{BEeta}
\end{equation}
where the sum runs over the fermions that are in equilibrium the thermal bath. Besides, $n_{\eta}^{eq}= T^{3}/\pi^{2}$ is the number density of a massless (real) scalar and $H\simeq 1.66\sqrt{g_{*}(T)}\,T^{2}/m_{\text{Pl}}$ is the expansion rate of the Universe. This equation is valid under the assumptions that  the SM fermions in Eq.~(\ref{BEeta}) are always in thermal equilibrium and that the Goldstone bosons remain in kinetic equilibrium right after the decoupling \cite{Gondolo:1990dk}, due to elastic scatterings. Moreover, we use the Boltzmann energy distribution for all the interacting particles, which is a good approximation for temperatures $T\lesssim 3\, m_{f}$, $m_{f}$ being the mass of the fermions produced in Goldstone boson annihilations.

The thermal averaged annihilation cross-section $\langle \sigma v \rangle_{\eta\eta\to f \bar{f}}$ is given by (see $e.g.$ \cite{Gondolo:1990dk}):
\begin{equation}
	\langle \sigma v \rangle_{\eta\eta\to f \bar{f}} \;\equiv\; \frac{1}{32\, T^{5}}\,\int_{4\,m_{f}^{2}}^{\infty}\,\sigma(\eta\eta\to f \bar{f})\,s\,\sqrt{s}\, K_{1}\left(\sqrt{s}/T\right)\,\text{d} s\,,\label{thermann}
\end{equation} 
with~\footnote{Notice that in the definition (\ref{thermann}) a factor 1/2 should be introduced to avoid double counting of the initial particle states. On the other hand,
the collision term in the Boltzmann equations (\ref{BEeta}) must be multiplied by 2 because of the annihilation of a pair of $\eta$'s. Therefore, 
the definitions (\ref{BEeta}) and (\ref{thermann}) are consistent.}
\begin{equation}
	\sigma(\eta\eta\to f \bar{f})=\frac{m_{f}^{2}\,\kappa^{2}}{8\pi}\,
	\,\frac{\left(1-4\,m_{f}^{2}/s\right)^{3/2}\,\left(s^{2}\left(m_{h}^{2}-m_{\rho}^{2}\right)^{2}\,+\,m_{\rho}^{2}\,m_{h}^{2}\,
	\left(m_{\rho}\,\Gamma_{h}-m_{h}\,\Gamma_{\rho}\right)^{2}\right)}{\left(m_{h}^{2}-m_{\rho}^{2}\right)^{2}\Big((s-m_{h}^{2})^{2}+\Gamma_{h}^{2}\,m_{h}^{2}\Big)\,\Big((s-m_{\rho}^{2})^{2}+\Gamma_{\rho}^{2}m_{\rho}^{2}\Big)}\,,\label{sigmaeta}
\end{equation}
where $\Gamma_{h,\rho}$ is the decay width of the scalar particle mediating the
s-channel annihilation cross-section.

The departure from equilibrium, in this case the  decoupling of $\eta$ from the plasma, according to Eq.~(\ref{BEeta}), takes place roughly at the temperature
 $T^{d}_{\eta}$ at which the following condition is satisfied:
 \begin{equation}
\frac{n_{\eta}^{eq}\;\sum_{f}\langle \sigma v \rangle_{\eta\eta\to f \bar{f}}}{H}\Bigg|_{T=T^{d}_{\eta}}\;=\;1\,\,.\label{deccond}
\end{equation}
\begin{figure}[t!]
\begin{center}
\includegraphics[width=14cm]{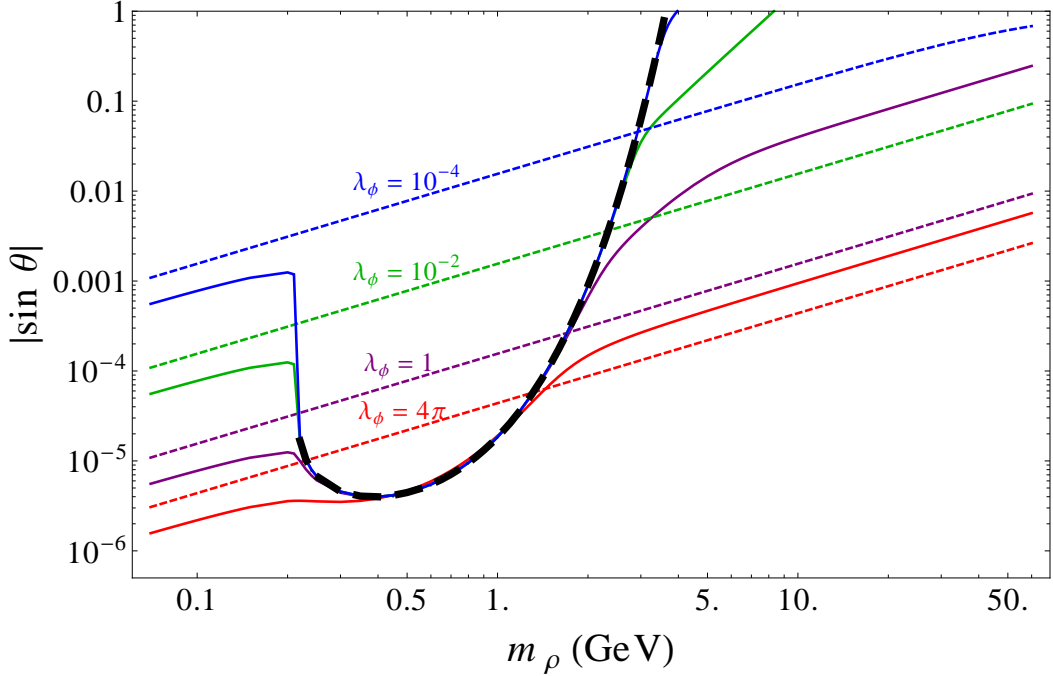}
\caption{Lower limit on $|\sin\theta|$ for fixed values of $\lambda_{\phi}$ (continuous lines) corresponding to Goldstone boson decoupling at $T\approx m_{\mu}$.
Upper limit on $|\sin\theta|$ for fixed values of $\lambda_{\phi}$ (dashed lines) given by the constraints on the Higgs boson invisible decay width. The thick dashed black curve 
is obtained using the analytic expression reported in Eq.~(\ref{Wcon2}).}
\label{figure:thetaLBDR}
\end{center}
\end{figure}
Using the previous expression, we can calculate the minimum value of $|\kappa|$ or, equivalently, 
$|\sin\theta|$ (see Eq.~(\ref{kappaeq})), for which
$\eta$ decouples form the thermal bath at temperature $T^{d}_{\eta}\approx m_{\mu}$. 
Notice that since the cross-section (\ref{sigmaeta}) is proportional to the squared mass of the fermion, it is enough to consider only the annihilation into  
$\mu^{\pm}$ pairs in (\ref{deccond}).

We report in Fig.~\ref{figure:thetaLBDR} the full numerical calculation of the lower limit of $|\sin\theta|$, for fixed values of the 
quartic coupling $\lambda_{\phi}$ (continuous lines). We also show the corresponding upper bound  derived from the invisible decay
width of the Higgs boson (dashed lines). We can see from this plot that there are three different regimes according to the value of the $\rho$ mass.
It turns out, that for each of them it is possible to find approximate analytical expressions.

\mathversion{bold}
\subsubsection*{Regime I: $m_{\rho}\gtrsim 4$ GeV}
\mathversion{normal}
In this regime, the values of the center of mass energy contributing to the integral are always much smaller the 
masses of the scalar particles which mediate the annihilation. As a result we can neglect $s$ in the denominator of
(\ref{sigmaeta}). Furthermore, if the decay width of the $\rho$ scalar can be neglected with respect to the other terms in
Eq.~(\ref{sigmaeta}), then the thermal annihilation cross-section into $\mu^{+}\,\mu^{-}$ in Eq.~(\ref{thermann}) is approximately:
\begin{equation}
	\langle \sigma v \rangle_{\eta\eta\to \mu^{+} \mu^{-}}\;=\; \frac{\kappa^{2}}{128\,\pi}\frac{m_{\mu}^{2}\,T^{4}}{m_{h}^{4}\,m_{\rho}^{4}}\,
	\int_{2\,m_{\mu}/T}^{\infty}\,w^{8}\,K_{1}(w)\,\text{d}w\,.
\end{equation}
Taking $T=T^{d}_{\eta}\simeq m_{\mu}$ in the previous equation, then the requirement of having a scalar
dark radiation component, Eq.~(\ref{deccond}),  implies
\begin{equation}
	|\kappa| \; \gtrsim \; \frac{m_{h}^{2}\,m_{\rho}^{2}}{m_{\text{Pl}}^{1/2}\, m_{\mu}^{7/2}}\,.\label{Wcon}
\end{equation}
This condition was derived for the first time in \cite{Weinberg:2013kea} 
and used to estimate a lower bound on $|\kappa|$ for $m_{\rho}\approx 500$ MeV.
Finally, from relation (\ref{kappaeq}) we can express the condition above as a lower limit on the scalar mixing angle $\theta$.
Namely, in this mass range for $\rho$ we have
\begin{eqnarray}
	|\sin\theta|\;\gtrsim\; 
	1.3\times 10^{-7}\,\lambda_{\phi}^{-1/2}\, \left(\frac{m_{\rho}}{0.1\,\text{GeV}}\right)^{3}\,.
\end{eqnarray}
Such values of the mixing angle are excluded by the present collider constraints on the invisible decay width of the Higgs boson derived previously, Eq.~(\ref{thetabound}).

In contrast, if the terms that depend on $\Gamma_{\rho}$ dominate in the numerator of (\ref{sigmaeta}), that is if $s<\Gamma_{\rho} \,m_{\rho}$ for the values of $s$ that
contribute to the integral, then the annihilation cross-section is described by
\begin{eqnarray}
	\langle \sigma v \rangle_{\eta\eta\to \mu^{+} \mu^{-}}&=& \frac{\kappa^{2}}{128\,\pi\,m_{h}^{2}}\left(\frac{m_{\mu\,}\Gamma_{\rho}}{m_{h}\,m_{\rho}}\right)^{2}\,
	\int_{2\,m_{\mu}/T}^{\infty}\,w^{4}\,K_{1}(w)\,\text{d}w\nonumber\\
	&=&\frac{\kappa^{2}}{128\,\pi\,m_{h}^{2}}\left(\frac{m_{\mu}}{m_{h}}\right)^{2}\,\left(\frac{\lambda_{\phi}}{16\pi}\right)^{2}
	\int_{2\,m_{\mu}/T}^{\infty}\,w^{4}\,K_{1}(w)\,\text{d}w\,,
\end{eqnarray}
where in the last term we replaced, at leading order in $\theta$, $\Gamma_{\rho}\simeq \lambda_{\phi}\, m_{\rho}/(16\pi)$.
We assume for simplicity $M_{\pm}\gtrsim 2$ GeV.
Notice that in this case,  the annihilation cross-section, for fixed $\lambda_{\phi}$, does not depend on $m_{\rho}$.

The corresponding lower limits in $|\kappa|$ and $|\sin\theta|$  now result:
\begin{equation}
	|\kappa| \; \gtrsim \; \frac{2\times 10^{3}}{\lambda_{\phi}}\frac{m_{h}^{2}}{m_{\text{Pl}}^{1/2}\, m_{\mu}^{3/2}}\,,
	\quad\quad\quad|\sin\theta|\;\gtrsim\;  2.8\times 10^{-4}\, \lambda_{\phi}^{-3/2}\left(\frac{m_{\rho}}{0.1\;\text{GeV}}\right)\,.\label{Wcon2}
\end{equation}
Combining the previous bound with the upper limit given by the invisible decay width of the Higgs, we get that the two bounds are consistent only for
a non-perturbative value of $\lambda_{\phi}$. Therefore we conclude that the Goldstone bosons cannot play the role of a dark radiation
for $m_{\rho}\gtrsim4$ GeV, regardless of the value of $\Gamma_{\rho}$.

\mathversion{bold}
\subsubsection*{Regime II: $2 \,m_{\mu} \lesssim m_{\rho}\lesssim 4$ GeV}
\mathversion{normal}
In this mass range the thermal annihilation cross-section is resonantly enhanced due to the fact that the annihilation
proceeds via s-channel (see Fig.~\ref{figure:etadecdiagrams}) and the typical center of mass energies contributing to the integral in (\ref{thermann}) are close to the 
$\rho$ mass. In particular, in the case of a narrow resonance, that is $(\Gamma_{\rho}/m_{\rho})^{2}\ll 1$, we can safely approximate
\begin{equation}
	\frac{1}{\pi}\,\frac{\Gamma_{\rho}\,m_{\rho}}{(s-m_{\rho}^{2})^{2}\,+\,\Gamma^{2}_{\rho}\,m_{\rho}^{2}}\;\to\;\delta(s-m_{\rho}^{2})\,.
\end{equation}
In this case the integral in (\ref{thermann}) is easily computed and we obtain an analytic expression of the averaged annihilation cross-section
in the given mass range:
\begin{eqnarray}
	\langle \sigma v \rangle_{\eta\eta\to \mu^{+} \mu^{-}}&=&
	\frac{\kappa^{2}}{256}\,\frac{m_{\mu}^{2}\,m_{\rho}^{6}}{T^{5}\,m_{h}^{4}\,\Gamma_{\rho}}\,
	\left(1-\frac{4\,m_{\mu}^{2}}{m_{\rho}^{2}}\right)^{3/2}\,K_{1}(m_{\rho}/T)\nonumber\\
	&=&\frac{\kappa^{2}\,\pi}{16}\,\frac{m_{\mu}^{2}\,m_{\rho}^{5}}{T^{5}\,m_{h}^{4}\,\lambda_{\phi}}\,
	\left(1-\frac{4\,m_{\mu}^{2}}{m_{\rho}^{2}}\right)^{3/2}\,K_{1}(m_{\rho}/T) \label{Wcon2}\,.
\end{eqnarray}
As we did above, we impose $T=T^{d}_{\eta}\simeq m_{\mu}$ and we derive the minimum value of $\kappa$
for which the Goldstone bosons may contribute to the effective number of relativistic neutrinos. Indeed, taking into account Eq.~(\ref{Wcon2}) we obtain
\begin{equation}
	|\kappa|\;\gtrsim\; 17\,\frac{m_{h}^{2}\,\lambda_{\phi}^{1/2}}{m_{\text{Pl}}^{1/2}\,m_{\mu}^{3/2}\,F(m_{\rho}/m_{\mu})}\;>\;
	5.3 \,\frac{m_{h}^{2}\,\lambda_{\phi}^{1/2}}{m_{\text{Pl}}^{1/2}\,m_{\mu}^{3/2}}\,\approx 7 \times10^{-4}\,\lambda_{\phi}^{1/2} \,,
\end{equation}
with $F(w)\equiv w\,(w^{2}-4)^{3/4}\,K_{1}(w)^{1/2}$. In the second inequality we report the least stringent bound, which corresponds to $m_{\rho}= 5.0\,m_{\mu}\simeq 525$ MeV,
where the function $F(m_{\rho}/m_{\mu})$ is maximized. 
The corresponding lower bound of $|\sin\theta|$ is independent of the quartic coupling $\lambda_{\phi}$  and is given by
\begin{equation}
	|\sin\theta|\;\gtrsim\; 17 \frac{v_{H}\,m_{\rho}}{\sqrt{2\,m_{\mu}\,m_{\text{Pl}}}\,F(m_{\rho}/m_{\mu})\,m_{\mu}}\;>\; 3\times 10^{-6}\,,\label{sinthetanalytic}
\end{equation}
where the minimum is obtained 
at $m_{\rho}=3.7\, m_{\mu}\simeq 390$ MeV. We report in Fig.~\ref{figure:thetaLBDR} the limit on $|\sin\theta|$ obtained using the analytic expression of the thermal annihilation cross-section given in Eq.~(\ref{Wcon2}) (thick dashed line). We can see that the analytic expression describes precisely the numerical lower bound.

\mathversion{bold}
\subsubsection*{Regime III: $m_{\mu}\lesssim m_{\rho}\lesssim 2\,m_{\mu}$}
\mathversion{normal}
In this case, at $T= T^{d}_{\eta}\approx m_{\mu}$ a fraction of the Goldstone bosons might  have enough kinetic energy to produce $\rho$ particles. Consequently, the latter are still present in the thermal bath and the Boltzmann equation (\ref{BEeta}), strictly speaking, is not appropriate to describe the $\eta$ decoupling. In order to account for that effect, the cross-section in Eq.~(\ref{deccond}) should also include annihilation of $\rho$ scalars into $\mu^{\pm}$ pairs.
Nevertheless, such process is phase space suppressed at these temperatures and as a result
we can still use  Eq.~(\ref{deccond}) to estimate a lower bound of the quartic coupling $\kappa$.
Under this assumption, the thermal annihilation cross-section is independent of $m_{\rho}$ and takes the  form:
\begin{eqnarray}
	\langle \sigma v \rangle_{\eta\eta\to \mu^{+} \mu^{-}}&\simeq&
	\frac{\kappa^{2}}{128 \,\pi}\,\frac{m_{\mu}^{2}}{m_{h}^{4}}\,\int_{2\,m_{\mu}/T}^{\infty}\,w^{4}\,K_{1}(w)\,\text{d}w\,,
\end{eqnarray}
and 
\begin{eqnarray}
	|\kappa|&\gtrsim& \frac{40\,m_{h}^{2}}{m_{\mu}^{3/2}\,m_{\text{Pl}}^{1/2}}\;\approx\; 5\times 10^{-3}\,,\\
	|\sin\theta|&\gtrsim &8\times 10^{-4}\, \left(\frac{v_{\phi}}{10\,\text{GeV}}\right)\;=\;5.7\times 10^{-6}\,\lambda_{\phi}^{-1/2}\, \left(\frac{m_{\rho}}{0.1\,\text{GeV}}\right)\,.
\end{eqnarray}
Notice that in this case the lower limit on $|\sin\theta|$ is compatible with the corresponding upper bound obtained from the invisible decay width of the Higgs boson, Eq.~(\ref{thetabound}).\\

\section{Constraining Dark Radiation with Direct Detection Experiments}
\label{sec:exclusionplots}

The results derived in Sections \ref{sec:direct_detection} and \ref{sec:dark_radiation} can be applied to find the regions of the parameter space that allow for Goldstone bosons as dark radiation and that are compatible with the negative searches of present dark matter direct detection experiments. On the one hand, the requirement of producing the correct $N_{eff}$, together with the requirement of perturbativity $\lambda_\phi < 4\pi$, gives a lower limit on $|\sin\theta|$ as a function of $m_\rho$, {\it cf.} Fig.~\ref{figure:thetaLBDR}. On the other hand, for thermally produced dark matter particles, the LUX experiment sets an upper limit on $|\sin\theta|$ as a function of $m_\rho$, as long as $m_{\rho}\lesssim 0.8\,M_{-}$, {\it cf.} Fig.~\ref{figure:sin2thetaBound}. Therefore, only some windows for $|\sin\theta|$ are allowed from the requirement of Goldstones as dark radiation and the non-observation of a signal at LUX. 

\begin{figure}[t]
\begin{center}
\includegraphics[width=12cm]{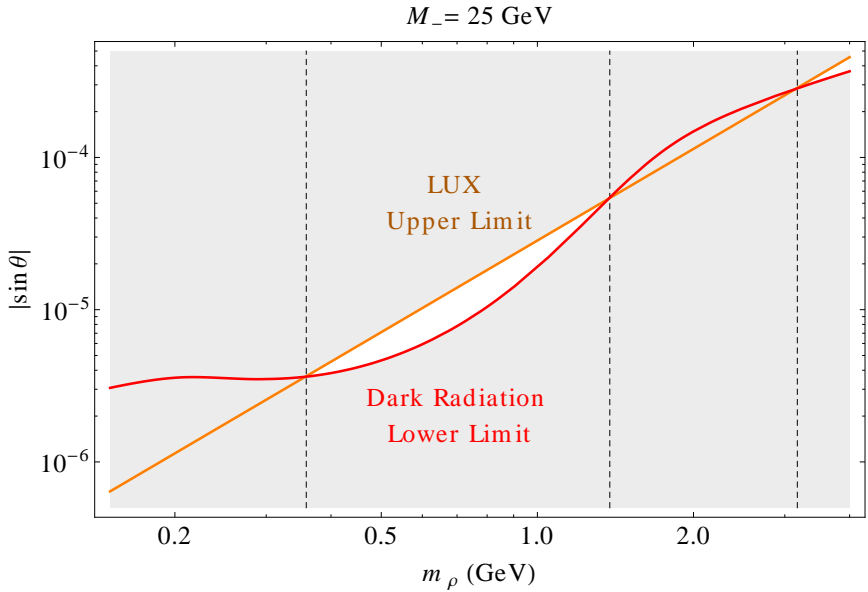} 
\caption{Allowed region  of $m_{\rho}$ (white areas) consistent with both the dark radiation hypothesis and the LUX limits for a 
dark matter mass of 25 GeV.}
\label{figure:example}
\end{center}
\end{figure}

\begin{figure}[t]
\begin{center}
\includegraphics[width=12cm]{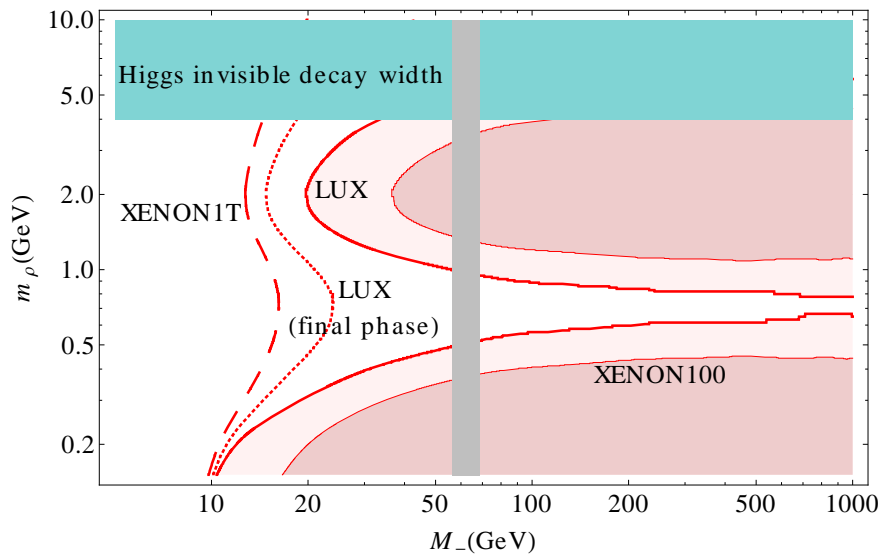} 
\caption{ Excluded regions of $m_{\rho}$ as a function of the dark matter mass under the hypothesis of Goldstone bosons as dark radiation.
The  dotted and dashed lines are the projected bounds from the final phase of the LUX and XENON1T experiments, respectively. We do not include in the analysis the
Higgs resonance region (gray band). }
\label{figure:ExclusionPlot}
\end{center}
\end{figure}

For illustration, we show this window in Fig.~\ref{figure:example} for a dark matter mass of $25 ~\GeV$, highlighting the values of $m_\rho$ where both limits coincide (dashed lines), which define the allowed  (white) and excluded regions (gray).
 By applying the same procedure we calculate the allowed regions of $m_{\rho}$ (red thick lines) for dark matter masses in the range $10 ~\GeV$ and $1000 ~\GeV$; the result is shown in Fig.~\ref{figure:ExclusionPlot}, where we shaded in light  red the areas excluded by LUX. For comparison we also show in dark red the regions excluded 
by the XENON100 experiment. 
  The cyan area,  $m_{\rho}>4$ GeV, corresponds to the regime I discussed in Section~\ref{sec:dark_radiation}, for which it is not possible to have dark radiation  due to the upper bound on  $\theta$  from the invisible Higgs decay width. Again, close to the Higgs boson resonance (gray band) the limits previously derived do not apply. For dark matter masses larger than $100$ GeV, dark radiation is allowed for $0.5~\GeV\lesssim m_{\rho}\lesssim 0.9~\GeV$. This case corresponds to the regime II for which Goldstone annihilation into $\mu^{\pm}$ pairs is  resonantly enhanced, thus allowing for $|\sin\theta|$ values that can evade the LUX bound, for any dark matter mass. This region is shown in the plot as the ``throat'' at $0.5~\GeV\lesssim m_{\rho}\lesssim 0.9~\GeV$. Besides, for $M_{-}\lesssim 19$ GeV the upper bound on $m_{\rho}$ given by direct detection disappears, making masses as large as $4$ GeV possible. 

 We also report in Fig.~\ref{figure:ExclusionPlot} the corresponding prospects for the direct detection experiments LUX (final phase)~\cite{Akerib:2012ys} and XENON1T~\cite{Aprile:2012zx}. It is remarkable that a large part of the parameter space will be probed by these two experiments. In the former case, dark matter masses larger than about 25 GeV could be excluded, whereas in the latter it would be possible to exclude masses even as low as 15 GeV.

\section{Conclusions}
\label{sec:conclusions}

We have studied a dark matter model where the stability of the dark matter particle is attributed to the residual $Z_2$ symmetry that arises from the spontaneous breaking of a global $U(1)$ symmetry. We have argued that the scalar sector responsible for the symmetry breaking plays a central role in the thermal production of  dark matter, since the dominant (co)annihilations channels at the time of freeze-out have in the final state the $CP$-even scalar of the dark sector and/or the Goldstone boson. We have carefully calculated the relic density and we have shown that, when the $CP$-even scalar of the dark sector is sufficiently light, there exists a lower limit on the dark matter coupling to the dark sector scalars from the requirement of reproducing the correct dark matter relic abundance. 

The dark sector in this model communicates with the Standard Model via the Higgs portal. As a result, after the spontaneous breaking of the electroweak symmetry and the dark global $U(1)$ symmetry, a mixing term arises between the Standard Model Higgs boson and the dark sector $CP$-even scalar. The Higgs phenomenology is accordingly modified and in particular the invisible Higgs decay width, since new decay channels into dark sector particles are possible. Besides, the mixing induces the scattering of dark matter particles with nucleons, thus opening the possibility of observing signatures of this model in direct dark matter search experiments. Working under the reasonable assumption that the dark matter particle was thermally produced in the early Universe, we have found a stringent upper limit on the mixing angle as a function of the mass of the dark $CP$-even scalar from the negative searches by the LUX experiment.

The massless Goldstone boson which is predicted by this model is, as recently remarked by Weinberg, an excellent candidate of dark radiation that could account for the exotic contribution to the number of effective neutrinos hinted by various experiments, provided the Goldstone bosons were in thermal equilibrium with the Standard Model particles until the era of muon annihilation. We have reevaluated,  using the exact expression of the thermal annihilation cross-section into $\mu^+\mu^-$, the values of the model parameters necessary to reproduce the central value of $N_{eff}$. We have then derived a lower limit on the mixing angle as a function of the mass of the dark $CP$-even scalar from the requirement of perturbativity of the quartic couplings of the model (evaluated at the weak scale). Lastly, we have combined the upper limit on the mixing angle which follows from the LUX experiment with the lower limit imposed by the requirement of dark radiation and we have found large regions of the parameter space where both requirements are incompatible. The final phase of LUX and the future XENON1T experiment will continue closing in on the parameter space of the model and will be able to rule out the possibility that the Goldstone boson contribute sizably to $N_{eff}$ if the dark matter mass is larger than $\sim 25$ GeV.

\section*{Acknowledgements}
We are grateful to Miguel Pato for useful discussions. 
This work was supported in part by the DFG cluster of excellence ``Origin and Structure
of the Universe'', by the ERC Advanced Grant project ``FLAVOUR''(267104) (A.I., E.M.) and by
the Graduiertenkolleg ``Particle Physics at the Energy Frontier of New Phenomena'' (C.G.C.).

\section{Appendix}

In this appendix we report the (co)annihilation cross-sections of $\psi_\pm$ for an arbitrary center of mass-energy in the limit $\theta=0$. 
We introduce for convenience the following notation
\begin{eqnarray}
\omega = \frac{\sqrt{s}}{M_-}\;,
\hspace{20pt}
u(r,w)= (z-1)^2\left(\left(\frac{\Gamma_\rho}{m_\rho}\right) ^2 r^4+\left(r^2-\omega ^2\right)^2\right)\,.
\end{eqnarray}
In terms of these variables, the cross-sections are given by 
\begin{align}
&\sigma (\psi_-\psi_-\to \eta\eta) (\omega) = \frac{f^4}{64 \pi  M_-^2  \omega ^2 \left(w^2-4\right) u(r,w) }\nonumber\\
&\Bigg[
\frac{\log \left(\frac{\omega ^2-\omega _1^2+z^2-2}{\omega _1^2+z^2}\right) }{ (z-1) \left(\omega ^2+2 z^2-2\right)}\Bigg(
4 r^4 \left(2 \omega ^2+4 z^4-2 \omega ^2 z^3+8 z^3+2 \omega ^2 z^2-\omega ^4 z+6 \omega ^2 z-8 z-4\right) 
\nonumber\\
&+4 r^2 \omega ^2
   \left(-2 \omega ^2-4 z^4+2 \omega ^2 z^3-8 z^3-2 \omega ^2 z^2+\omega ^4 z-6 \omega ^2 z+8 z+4\right)
\nonumber\\
&+u(r,w) (z-1) \left(\omega ^4+6 z^4+8 z^3+4
   \left(2 \omega ^2-3\right) z^2+8 \left(\omega ^2-3\right) z-10\right)
\Bigg)
\nonumber\\
&
+\frac{4 \omega _1^2
   \left(-r^4 \omega ^2+4 r^4 z+4 r^2 \omega ^2-4 r^2 \omega ^2 z+u(r,w) z^2-2 u(r,w) z+u(r,w)\right)}{(z-1)^2}
\nonumber\\
&
-\frac{(z+1)^4 u(r,w)}{\omega_1^2-\omega ^2-z^2+2}
-\frac{(z+1)^4 u(r,w)}{\omega _1^2+z^2}
\Bigg]^{\omega_1 = \sqrt{\frac{w^2}{4}-1}- \frac{w}{2} }_{\omega_1 = \sqrt{\frac{w^2}{4}-1}+  \frac{w}{2} }\,,
\end{align}

\begin{align}
&\sigma (\psi_-\psi_-\to \rho\rho) (\omega) = \frac{f^4}{64 \pi M^2_-   \omega ^2 \left(w^2-4\right) }\nonumber\\
&\Bigg[
\frac{\log \left(\frac{\omega _1^2+1}{2 r^2-\omega ^2+\omega _1^2+1}\right)}{(z-1) \left(2 r^4-3 r^2 \omega ^2+\omega
   ^4\right)}\Bigg( 6 r^6 (z+7)-2 r^4 \left(-5 \omega ^2+5 \omega ^2 z+8 z+88\right) \nonumber\\
& +r^2 \left(-17 \omega ^4+64 \omega ^2+5 \omega ^4 z+32 \omega
   ^2 z-32 z+32\right)+\omega ^2 \left(\omega ^4+16 \omega ^2-32\right) (1-z)\Bigg)\nonumber\\
&+\frac{4 \omega _1^2 \left(-3
   r^2 \omega +r^2 z+5 r^2+\omega ^2-\omega ^2 z\right) \left(3 r^2 \omega +r^2 z+5 r^2+\omega ^2-\omega ^2 z\right)}{(z-1)^2 \left(r^2-\omega
   ^2\right)^2} \nonumber\\
&-\frac{\left(r^2-4\right)^2}{2 r^2-\omega ^2+\omega _1^2+1}-\frac{\left(r^2-4\right)^2}{\omega _1^2+1}
\Bigg]^{\omega_1 = \sqrt{\frac{\omega^2}{4}-1}- \sqrt{\frac{\omega^2}{4}-r^2} }_{\omega_1 = \sqrt{\frac{\omega^2}{4}-1}+ \sqrt{\frac{\omega^2}{4}-r^2} }\,,
\end{align}

\begin{align}
&\sigma (\psi_-\psi_+\to \rho\eta) (\omega) = \frac{f^4}{64 \pi  M_-^2\Big(\omega ^2-(z-1)^2\Big) \Big(\omega ^2-(z+1)^2\Big)}\nonumber\\
&\Bigg[
-\frac{4r^2 (z+1) \left( r^2+\omega ^2-2 z^2-4 z-2\right) }{\omega ^2 (z-1)} \log \left[\left(1-\omega _1^2\right) \left(r^2-\omega ^2-\omega _1^2+1\right)\right]\nonumber\\
&-\frac{2  \log \left(\frac{r^2-\omega ^2-\omega _1^2+1}{1-\omega _1^2}\right)}{\omega ^2 (\omega^2-r^2 )} \Big(2 r^6-5 r^4 \omega ^2+4 r^4 z^2 +8 r^4 z+4 r^4+4 r^2 \omega ^4-2 r^2 \omega ^2-2 r^2 \omega ^2 z^2\nonumber\\
&-4 r^2 \omega ^2 z-\omega ^6-8 \omega ^2 z^3-16 \omega ^2 z^2-8 \omega ^2
   z\Big)+\frac{8 r^4 \omega _1^2 (\omega^2 - (z+1)^2) }{\omega ^4 (z-1)^2}-\frac{2 (z+1)^2
   (r^2-4 z^2)}{\omega _1^2-r^2+\omega ^2-1}\nonumber\\
&-\frac{2 \left(r^2-4\right) (z+1)^2}{\omega _1^2-1}
\Bigg]^{\omega_1^2 = \frac{r^2 \omega ^2+r^2 z^2-r^2-\omega ^4+\omega
   ^2+\omega ^2 z^2+\left(\omega ^2-r^2\right) \sqrt{\left(\omega ^2-1\right)^2+z^4-2 \left(\omega ^2+1\right) z^2}}{2 \omega ^2}}_{\omega_1^2 = \frac{r^2 \omega ^2+r^2 z^2-r^2-\omega ^4+\omega
   ^2+\omega ^2 z^2-\left(\omega ^2-r^2\right) \sqrt{\left(\omega ^2-1\right)^2+z^4-2 \left(\omega ^2+1\right) z^2}}{2 \omega ^2}}\,,
\end{align}
\begin{equation}
\sigma (\psi_+\psi_+\to \eta\eta) (\omega)= \frac{1}{z^2}\sigma (\psi_-\psi_-\to \eta\eta)\left(\frac{\omega}{z}\right)\Bigg|_{r\to \frac{r}{z},z\to \frac{1}{z}}\,, \hspace{5cm}
\end{equation}
\begin{equation}
\sigma (\psi_+\psi_+\to \rho\rho) (\omega)= \frac{1}{z^2}\sigma (\psi_-\psi_-\to \rho\rho)\left(\frac{\omega}{z}\right)\Bigg|_{r\to \frac{r}{z},z\to \frac{1}{z}}\,. \hspace{5cm}
\end{equation}

\end{document}